\documentclass[ aps, showpacs, showkeys, nofootinbib, floatfix, superscriptaddress]{revtex4}

\usepackage{amsfonts}

\usepackage{amssymb}
\usepackage{amsmath}
\usepackage{graphicx}
\usepackage{epstopdf}
\usepackage{multirow}
\usepackage[utf8]{inputenc}

\begin{document}
\title{Constraining magnetized black-bounce spacetime from HFQPOs data and selecting resonance models via information criteria and Bayesian analysis}
 \author{Shining Yang}
 \affiliation{Department of Physics, Liaoning Normal University, Dalian 116029, P. R. China, Shanghai Normal University Tianhua College, Shanghai 201815, P. R. China}
 \author{Jianbo Lu}
 \email{lvjianbo819@163.com}
 \affiliation{Department of Physics, Liaoning Normal University, Dalian 116029, P. R. China}
 \author{Mou Xu}
 \affiliation{Department of Physics, Liaoning Normal University, Dalian 116029, P. R. China}
 \author{Yu Liu}
 \affiliation{Department of Physics, Liaoning Normal University, Dalian 116029, P. R. China}

\begin{abstract}
Black bounce spacetime, proposed by Simpson and Visser (SV), can describe the Schwarzschild solution, regular black holes, and traversable wormholes in a unified manner, depending on the value of the regularization parameter $a$. This paper primarily explores the dynamics of charged particle in the magnetized SV spacetime, and constrains the parameters of the SV spacetime along with its surrounding magnetic fields. The constraints are given by using $\chi^2$ analysis combined with high-frequency quasi-periodic oscillation (HFQPO) data observed from three microquasars: GRS 1915+105, XTE 1550-564, and GRO 1655-40. The results indicate that the magnetic field significantly influences the position of the innermost stable circular orbit of charged particle and frequency distribution of epicyclic motion, which excites more resonance model variants, enhancing observational effects. Additionally, we employ the Akaike Information Criterion (AIC) to evaluate  resonance model and its various variants. The support for different models from observational data shows significant variation: $E R_8$ as the best model is supported  strongly, $ER_3$ model has moderate evidence of support,  $ER_6$ and $ER_7$ models are considerably less support, while other resonance models have essentially no support. The result from AIC is broadly consistent with the results obtained from the Bayesian Information Criterion (BIC) and the Bayes factor. For models more supported by the observational data, the allowed ranges of the regularization parameter: $0\leq a/M<0.736$ ($68\%$ confidence level) suggests that HFQPOs data could support the magnetized black bounce spacetime as a regular black hole, and the smaller value of the regularization parameter indicates a possibility of the presence of quantum effects. According to the constraint results, we get the best-fit values of magnetic field strength around $10^{-5}\sim 10^{-4}$ GS for electrons and around  $10^{-2}\sim 10^{-1}$ GS for protons. Finally, as a comparison, we test the SV spacetime without a magnetic field using microquasar observational data, and the calculated results of AIC and BIC show that this case is incompatible with the HFQPOs data, further supporting the existence of a magnetic field in SV spacetime.

\end{abstract}


\maketitle

\section{$\text{Introduction}$}
In recent years, substantial advancements have been made in the field of black hole physics. Empirical observations, such as the detection of gravitational waves \cite{1} and the imaging of black hole shadows \cite{2,3,wei} have provided strong validation for the theoretical predictions of General Relativity (GR). Nonetheless, the occurrence of unavoidable spacetime singularities within GR points to the breakdown of classical physical laws in extreme gravitational fields. Although quantum gravity is widely considered a potential solution to this issue, the theory remains incomplete. Consequently, the exploration of phenomenological models that may account for quantum gravitational effects has gained considerable academic interest. Several such models have been proposed and extensively studied, including singularity-free gravitational collapse \cite{4,5,6,7,8}, nonsingular cosmological models \cite{9,10,11,12,13,14,15}, and regular black hole solutions \cite{16,17,18,19,20,21,22,23,24}.

Among the proposed modifications to classical black holes, the concept of a regular black hole was first introduced by Bardeen in 1968 \cite{16}. Building upon this foundation, Simpson and Visser (SV) later developed a spacetime metric referred to as the "black bounce" \cite{25}. This metric provides a unified framework that encompasses the Schwarzschild solution, regular black holes, and traversable wormholes, offering a simplified representation of quantum gravitational effects \cite{26}. Extensive research has since explored various aspects of the SV spacetime, including its gravitational lensing properties, quasi-periodic oscillations (QPOs), accretion disks, shadows, and quasi-normal modes, etc \cite{26,27,28,29,30,31,32,33,34,35,36,37,38,39,40,42}.

QPOs, due to their high precision, have emerged as a powerful tool for testing gravitational theories \cite{43,44,45,46,47,48,49,50}. These oscillations manifest as peaks in the electromagnetic spectrum, spanning from radio to X-ray frequencies, and are associated with compact objects \cite{51}. QPOs are generally classified into two categories: Low Frequency (LF) and High Frequency (HF), depending on the observed oscillation frequencies. Through spectral analysis of QPOs \cite{52,53,54,56}, researchers can extract specific physical insights regarding the central compact object. However, the precise origin of QPOs remains elusive. It is widely hypothesized that they arise from precession and resonance phenomena linked to the effects of GR \cite{57,58,59}. In response, various theoretical models have been proposed, which can broadly be categorized into four types: the epicyclic resonance (ER) model \cite{60,61}, the relativistic precession (RP) model \cite{62,63}, the tidal disruption (TD) model \cite{64,65,66}, and the warped disc (WD) model \cite{67}. These models typically involve linear combinations of the radial, latitudinal, and orbital frequencies of particle in epicyclic motion around a central object \cite{65,69}. Despite significant efforts, no current theoretical models can fully reconcile observational data across different sources, and the exact physical mechanism underlying HFQPOs remains unresolved \cite{53,70}. It has been proposed that considering the epicyclic motion of charged particle in the vicinity of a black hole immersed in a magnetic field may offer new insights into this problem \cite{72,73,74}.

To date, many black hole candidates have been observed to possess accretion disks composed of plasma, whose dynamics can generate magnetic fields. Another potential source of electromagnetic fields around black holes is the amplification of external galactic magnetic fields by the strong gravitational pull of the black hole. This suggests that black holes may be immersed in large-scale external magnetic fields. These fields could possess globally complex structures; however, when the distance between the black hole and the magnetic field source is sufficiently large, the magnetic field within a finite spatial region around the black hole can be approximated as locally uniform. This locally uniform, weak magnetic field can be described  by the Wald solution \cite{76,77}. It is important to note that such weak fields do not significantly alter the geometric structure of the black hole itself \cite{78,79,80,81,82,83,85}. Nevertheless, even a small magnetic field B, due to the effects of the Lorentz force, can have a profound influence on the dynamics of charged particle \cite{86,87,88,89,90,91,92,93,94,95}. Thus, it is essential to consider the possible impacts of electromagnetic fields when studying black holes. The motion of charged test particle in asymptotically uniform magnetic fields around compact objects has been extensively studied, contributing significantly to our understanding of astrophysical processes near magnetized black holes \cite{96,97,98,99,100,102,103,105,106}. In this paper, we examine the effects of an external asymptotically uniform magnetic field on the dynamics of charged particle in SV spacetime and compare the theoretical results with observed HFQPO frequencies in three types of microquasars.

The structure of this article is as follows. The first part is the introduction. The second part of the article briefly introduces the SV spacetime, which can describe some typical black holes and traversable wormholes in a unified way. In the third and fourth sections, we show a uniform magnetic field and study the dynamical behavior of a moving particle in the equatorial plane of the SV spacetime immersed in the uniform magnetic field. We provide the location of the innermost stable circular orbit (ISCO) and derive the general expressions for the radial and latitudinal angular frequencies of the oscillating particle in its epicycle motion. In the fifth section, we introduce the resonance model and its variant forms, showing the locations where particle oscillating near the stable circular orbit undergo resonance under different models. So far, a large number of HFQPO models have been proposed in the literature, and the study of HFQPOs related theories largely depends on selected model. However, none of these models have gained widespread acceptance (such as in observations). So, it is very meaningful to evaluate various HFQPO models through statistical analysis of observational datasets. Among them, the double peak 3:2 frequency ratio observed in three Galactic microquasars (GRS 1915+105, XTE 1550-564, and GRO 1655-40) confirms that the nonlinear resonance between two modes of oscillation in the accretion disk around compact objects plays a role in exciting the observed X-ray flux modulation \cite{60}. Therefore, in section VI, we compare the quasi-harmonic oscillation frequencies of charged test particle under resonance model and its variants with the HFQPO data observed in the aforementioned three specific microquasars. We also apply these observational data to study the Akaike Information Criterion (AIC), Bayesian Information Criterion (BIC), and Bayes factor for model selection while constraining the regularization parameter and magnetic parameters in the magnetized SV spacetime. Finally, as a comparison, we also fit the observational data to constrain the SV spacetime without a magnetic field.

\section{$\text{The black bounce metric}$}

 Black bounce spacetime was proposed by Simpson and Visser, which describes a static, spherically symmetric line element \cite{25}:
\begin{equation}
\mathrm{ds}^2=-\mathrm{f}(\mathrm{r}) \mathrm{dt}^2+\mathrm{f}^{-1}(\mathrm{r}) \mathrm{dr}^2+\mathrm{h}(\mathrm{r})\left(\mathrm{d} \theta^2+\sin ^2 \theta \mathrm{d} \phi^2\right), \label{1}
\end{equation}
\begin{equation}
\mathrm{f}(\mathrm{r})=1-\frac{2 \mathrm{M}}{\sqrt{a^2+r^2}}, \quad \mathrm{~h}(\mathrm{r}) \equiv a^2+r^2, \label{2}
\end{equation}
where M is the mass of central body and the parameter $a$ in charge of the regularization of the central singularity with $a>0$. Under the influence of parameters $a$, several types of spacetime can be described by the SV geometry, the Schwarzschild spacetime $(a=0)$, a regular black hole $(0<a<2 \mathrm{M})$, a one-way wormhole with a null throat $(a=2 \mathrm{M})$ and a traversable wormhole $(a>2 \mathrm{M})$. In the subsequent calculations herein, we adopt the unit $\mathrm{M}=1$.

The source of SV metric is a combination of a minimally coupled phantom scalar field and a nonlinear electrodynamics field in the framework of GR \cite{108}. The action is:
\begin{equation}
S=\int \sqrt{-\operatorname{g}} \mathrm{d^4x}\left(R+2 \epsilon \mathrm{g}^{\mu \nu} \partial_\mu \phi \partial_\nu \phi-2 \mathrm{~V}(\phi)-\mathcal{L}(\mathcal{F})\right), \label{3}
\end{equation}
with $\epsilon=-1$, and $\mathcal{L}(\mathcal{F})$ is the Lagrangian density of a nonlinear electromagnetic field $\mathcal{F}=\mathcal{F}_{\mu \nu} \mathcal{F}^{\mu \nu}$. $R$ is the Ricci scalar, and $g$ denotes the determinate of metric $g_{\mu \nu}$. The forms of $\mathcal{L}(\mathcal{F})$  and the potential of a phantom scalar field $\phi(\mathrm{r})$ are \cite{108}:
\begin{equation}
\mathcal{L}(\mathcal{F})=\frac{12 \mathrm{M}a^2}{5\left(2 \mathrm{q}^2 / \mathcal{F}\right)^{5 / 4}}, \label{4}
\end{equation}
\begin{equation}
\mathrm{V}(\phi)=\frac{4 \cos ^6 \phi}{5 a^3} \operatorname{Msec}(\phi). \label{5}
\end{equation}

In addition, we notice that the phantom field, as a significant candidate for dark energy with an equation of state $\omega<-1$, has been widely explored to explain the late-time accelerated expansion of the universe. Additionally, in string theory, the phantom field manifests as a negative tension brane, playing a crucial role in string dualities \cite{108,109,110}. Within the framework of GR, the formation of a wormhole typically requires the presence of a large amount of exotic matter that violates the null energy condition, such as the phantom scalar field. To date, numerous wormhole solutions involving various types of phantom matter have been proposed \cite{108,111,112,113,114}. Meanwhile, some black hole solutions receiving from gravity coupled with phantom scalar fields or phantom Maxwell fields have been identified, and extensive research has been conducted on their associated geometric structures and thermodynamic properties \cite{115,116,117,118,119,120,121,122,123,124,125}.

It should also be noted that although phantom fields remain a theoretically interesting possibility, there are open issues regarding phantom models in both observations and theory. For instance, Godłowski and Szydłowski showed that a phantom model can provide a better fit to supernova observations than the $\Lambda\mathrm{CDM}$ model only if the present-day matter density parameter satisfies $\Omega_{0m}>0.34$ \cite{151}. This requirement seems to be inconsistent with some other current cosmic observational constraints on the matter density, e.g. the Planck 2018 CMB results yield $\Omega_{0m}=0.315\pm0.007$ ($68\%$ CL) \cite{PlanckCollaboration2020}, while analyses based on DESI BAO measurements likewise favor a lower matter density, $\Omega_{0m}=0.295\pm0.015$ ($68\%$ CL) \cite{DESI2025}. In addition, the theoretical self-consistency of phantom models remains controversial, primarily because they involve a kinetic term with the wrong sign and are generally associated with violation of the null energy condition \cite{Caldwell2002,Carroll2003}.

 \section{$\text{Dynamics of charged particle in a magnetized SV spacetime}$}

Now let us assume that there is an external asymptotically uniform magnetic field surrounding the SV space-time, and the magnetic field lines is perpendicular to the equatorial plane of the central celestial body, in which strength is $B_0$. Close to the strong gravitational field, the nonzero component of the 4-vector potential $A^\mu$ and the electromagnetic field tensor $A_\phi$ have the following form \cite{126,127,128}:
\begin{equation}
A^\mu=\frac{1}{2}\left(0,0, B_0, 0\right), \label{6}
\end{equation}
\begin{equation}
A_\phi=\frac{B_0}{2} g_{\phi \phi}=\frac{B_0}{2} r^2 \sin ^2 \theta. \label{7}
\end{equation}

To obtain the equation of motion of charged particle in the magnetized SV spacetime, we consider the Hamilton-Jacobi equation:
\begin{equation}
g^{\alpha \beta}\left(\frac{\partial S}{\partial x^\alpha}+q A_\alpha\right)\left(\frac{\partial S}{\partial x^\beta}+q A_\beta\right)=-m^2, \label{8}
\end{equation}
where $m$ and $q$ represent the mass and charge of the test particle, respectively. The symmetry of SV spacetime will not be destroyed by external uniform magnetic field. Due to the existence of timelike $\xi_{(t)}^\mu$ and spacelike $\xi_{(\phi)}^\mu$ Killing vectors, we can express the action of the test particle as:
\begin{equation}
S=-E t+L \phi+S_{r \theta}(r, \theta), \label{9}
\end{equation}
where both $E$ and $L$ are conserved quantities, representing the energy and angular momentum of the test particle, respectively. And they can be represented as:
\begin{equation}
E=-\xi_{(t)}^\mu\left(m v_\mu+q A_\mu\right)=m f(r) \frac{d t}{d \tau}, \label{10}
\end{equation}
\begin{equation}
L=\xi_{(\phi)}^\mu\left(m v_\mu+q A_\mu\right)=m\left(a^2+r^2\right) \sin \theta^2\left(\frac{d \phi}{d \tau}+\frac{q B}{2 m}\right), \label{11}
\end{equation}
here $v^\mu$ is the four-velocity of the test particle, $\tau$ is the affine parameter. To simplify calculation, the following relationship is established with considering a test particle of unit mass \cite{86}:
\begin{equation}
\varepsilon=\frac{E}{m} \quad, \quad \mathcal{B}=\frac{q B}{2 m}, \quad \mathcal{L}=\frac{L}{m}. \label{12}
\end{equation}

Utilizing the normalization condition of the four-dimensional velocity $g_{\mu \nu} v^\mu v^\nu=-1$, according to equations (\ref{10})-(\ref{12}) we obtain the motion equation of charged particle:
\begin{equation}
\frac{d t}{d \tau}=\frac{\varepsilon}{f(r)}, \label{13}
\end{equation}
\begin{equation}
\left(\frac{d r}{d \tau}\right)^2=\varepsilon^2-V_{e f f}, \label{14}
\end{equation}
with
\begin{equation}
V_{e f f}=f(r)\left[1+\left(\frac{\mathcal{L}}{\sqrt{h(r)} \sin \theta}-\sqrt{h(r)} \mathcal{B} \sin \theta\right)^2\right]. \label{15}
\end{equation}
In the parentheses of equation (\ref{15}), the specific angular momentum $\mathcal{L}$ and the magnetic parameter $\mathcal{B}$ give separately the central force potential and the electromagnetic potential energy \cite{131}.

It is easy to see from equation (\ref{15}) that the effective potential exhibits an obvious symmetry between $(\mathcal{L}, \mathcal{B})$ and $(-\mathcal{L},-\mathcal{B})$. It indicates that when one  changes the symbol of $(\mathcal{L}, \mathcal{B})$ to $(-\mathcal{L},-\mathcal{B})$, the effective potential will not change \cite{86,131}. If we fix the axisymmetric direction to be upward along the polar direction, then a positive value of angular momentum $(\mathcal{L}>0)$ hints that the particle will revolve in a counter-clockwise motion around the SV celestial body. Meanwhile, for the test particle with $\mathrm{q}>0$ and $\mathcal{B}>0$, the magnetic field direction is upward along the polar direction, while $\mathcal{B}<0$ is opposite.

Next, we consider the motion of charged test particle around the equatorial plane of the SV celestial body $(\theta=\pi / 2)$ under the influence of an external uniform magnetic field.  Motion of particle on circular orbit need to satisfy:
\begin{equation}
d V_{e f f}(r) / d r=0. \label{16}
\end{equation}
Through formulas (\ref{15}) and (\ref{16}) we obtain:
\begin{equation}
2 r\left\{r^2-2 \mathcal{B} \mathcal{L} r^2-\mathcal{B}^2 r^4+\mathcal{B}^2 r^4 x-\mathcal{L}^2(-3+x)+a^4 \mathcal{B}^2(-1+x)+a^2\left[2 \mathcal{B} \mathcal{L}+2 \mathcal{B}^2 \mathcal{L}^2(-1+x)\right]\right\}=0, \label{17}
\end{equation}
where $x=\sqrt{r^2+a^2}$, the energy of particle moving on a circular orbit is expressed as  $\varepsilon^2=V_{e f f}$. As a function of the radial coordinate $r$, the real root of equation (\ref{17}) represents the location of the stable or unstable circular orbit of the test particle. Due to the presence of higher-order terms containing $r$, it is difficult to obtain a corresponding analytical solution for equation (\ref{17}). However, it is quadratic in terms of a specific angular momentum $\mathcal{L}$, so the circular orbit can be determined by the following relationship:
\begin{equation}
\mathcal{L}_{e \pm}=\frac{-\mathcal{B} x^2 \pm \sqrt{x^2\left[-3+x+\mathcal{B}^2(-2+x)^2 x^2\right]}}{-3+x}. \label{18}
\end{equation}
The solution of $d \mathcal{L}_{e \pm} / d r=0$ corresponds to the location of the ISCO, which can be obtained by solving $d^2 V_{e f f}(r) / d r^2=0$:
\begin{equation}
\begin{aligned}
&\begin{aligned}
& \mathcal{L}_{e x}=\frac{x}{-a^2-2 r^2+3 x}\left\{a^2 \mathcal{B}+3 \mathcal{B} r^2-\frac{1}{2}\left[4 \mathcal{B}^2\left(a^2+3 r^2\right)^2-\frac{4}{x}\left[a^4 \mathcal{B}^2+r^2\left(-3+\mathcal{B}^2\left(5 r^2-\right.\right.\right.\right.\right. \\
& \left.\left.\left.6 x))+a^2\left(-1+\mathcal{B}^2\left(6 r^2-x\right)\right)\right]\left(a^2+2 r^2-3 x\right)\right]^{1 / 2}\right\}
\end{aligned}. \label{19}
\end{aligned}
\end{equation}

\begin{figure}[ht]
\includegraphics[width=9.8cm]{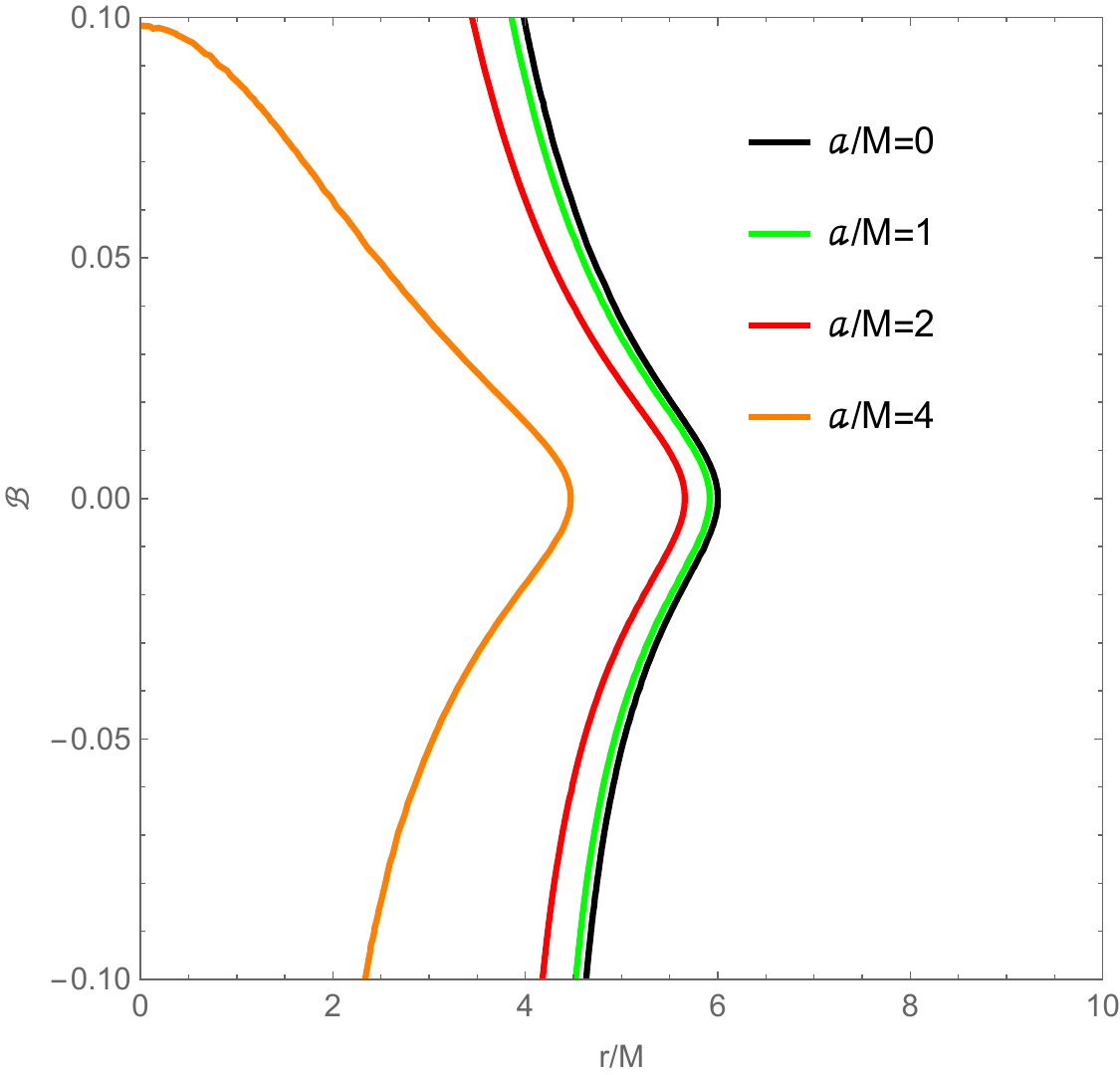}
  \caption{The variation of the ISCO position of a charged particle  with the magnetic field, where the particle is considered as  moving in SV spacetime around different types of celestial bodies.}
    \label{fig:1}
\end{figure}

Since the ISCO is obtained through solving $d V_{e f f}(r) / d r=0$ and $d^2 V_{e f f}(r)/d r^2=0$, it is located at the intersection of $\mathcal{L}_{e \pm}$ and $\mathcal{L}_{\text {ex}}$. In Figure \ref{fig:1}, we show how the position of the ISCO of a charged particle in a magnetized SV spacetime varies  with the magnetic field around different types of celestial bodies, e.g. the Schwarzschild BH $(a / M=0)$, a regular BH $(a / M=$ 1), a one-way BH with a null throat $(a / M=2)$ and a traversable wormhole $(a / M=$ 4).

From Figure \ref{fig:1}, we can see that when the magnetic field strength is fixed, the ISCO position of the charged particle moves closer to the radial coordinate center as the regularization parameter $a / M$ increases. This indicates that near the SV wormhole, the ISCO is closer to the radial coordinate center compared to the SV black hole. Additionally, for the same type of SV celestial body, the ISCO of the charged particle moves closer to the radial coordinate center as the magnetic field strength $|\mathcal{B}|$ increases, and the largest $r_{ISCO}$ appears for the case without a magnetic field. And, relatively speaking, for the same value $|\mathcal{B}|$, the ISCO radius for $\mathcal{B}>0$ is always smaller than that for $\mathcal{B}<0$. Hence, for SV black holes $(a/M=0, 1, 2)$, the presence of a uniform magnetic field results the circular orbit of the charged particle around the central celestial body to move closer to the event horizon. In contrast, for SV wormholes $(a / M=4)$, under the influence of a uniform magnetic field, the particle's trajectory can approach the surface of the wormhole infinitely, which further aids in receiving relevant information about the central celestial body.

\section{$\text{The angular frequency of the motion of a charged particle in a magnetized SV spacetime}$}
In this section, we study the oscillatory epicyclic motion of test particle in the magnetized SV spacetime. For a test particle moving in a circular orbit at position $r_c$ in the equatorial plane $\theta=\pi / 2$ of the central celestial body, if it experiences perturbations with small displacements $\delta r$ and $\delta \theta$ relative to $r_c$ and $\theta=\pi / 2$, the particle will oscillate. Within the range of linear perturbation, this motion is considered harmonic, and the equations governing the particle's epicyclic motion around a stable circular orbit in the radial and latitudinal directions may be represented as follows:
\begin{equation}
\delta \ddot{r}+\omega_r^2 \delta r=0, \quad \delta \ddot{\theta}+\omega_\theta^2 \delta \theta=0 . \label{20}
\end{equation}
Here, "dot" represents the derivative with respect to the particle's proper time $\tau$, and $\omega_{\mathrm{r}}$ (or $\omega_\theta$) denotes the angular frequency of the particle's radial (or latitudinal) oscillatory motion at the position of circular orbit. The expressions for both are given by the following equations \cite{86,128}:
\begin{equation}
\omega_r^2=\frac{\partial^2 V_{e f f}}{\partial r^2}, \label{21}
\end{equation}
\begin{equation}
\omega_\theta^2=\frac{1}{h(r) f(r)} \frac{\partial^2 V_{e f f}}{\partial \theta^2}. \label{22}
\end{equation}

Using equations (\ref{1}), (\ref{2}), and (\ref{15}), the expressions for $\omega_r^2$ and $\omega_\theta^2$ are  further derived as follows:
\begin{equation}
\begin{aligned}
& \omega_r^2=\frac{2}{x^7}\left\{-2 r^4+4 \mathcal{B} \mathcal{L} r^4-a^2\left[r^2-2 \mathcal{B} \mathcal{L} r^2+\mathcal{B}^2 \mathcal{L}^4(1-3 x)+\mathcal{L}^2(-3+x)\right]+\right. \\
& \left.3 \mathcal{L}^2 r^2(-4+x)+a^6 \mathcal{B}^2(-1+x)+\mathcal{B}^2 r^6 x+a^4\left[1-2 \mathcal{B} \mathcal{L}+\mathcal{B}^2 r^2(-2+3 x)\right]\right\}
\end{aligned}, \label{23}
\end{equation}
\begin{equation}
\omega_\theta^2=-\mathcal{B}^2+\frac{\mathcal{L}^2}{x^4}. \label{24}
\end{equation}
Furthermore, the orbital frequency $\omega_\phi$ of the particle's motion is expressed as:
\begin{equation}
\omega_\phi=\frac{d \phi}{d \tau}=\frac{\mathcal{L}}{r^2+a^2}-\mathcal{B}, \label{25}
\end{equation}
where $\mathcal{L}=\mathcal{L}_{+}$. In addition, the Larmor angular frequency $\omega_L$ caused by the uniform magnetic field itself is written as:
\begin{equation}
\omega_L=\frac{B q}{m}=2|\mathcal{B}|. \label{26}
\end{equation}

\begin{figure}[ht]
\includegraphics[width=4.8cm]{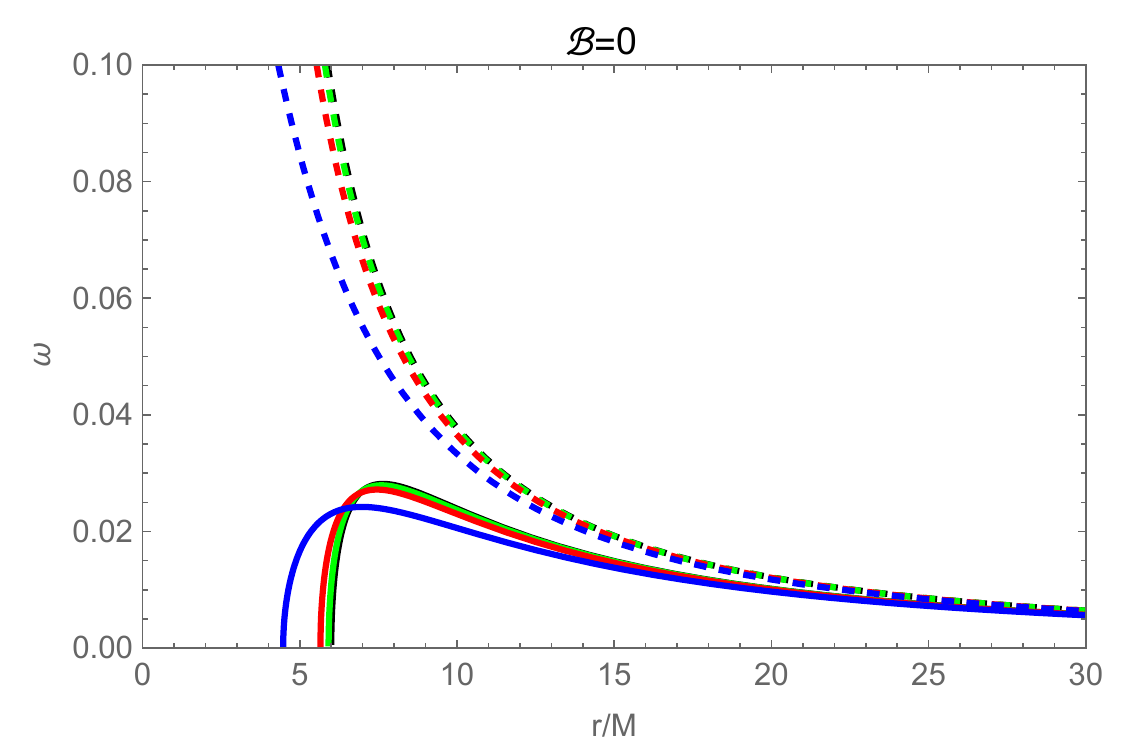}
\includegraphics[width=4.8cm]{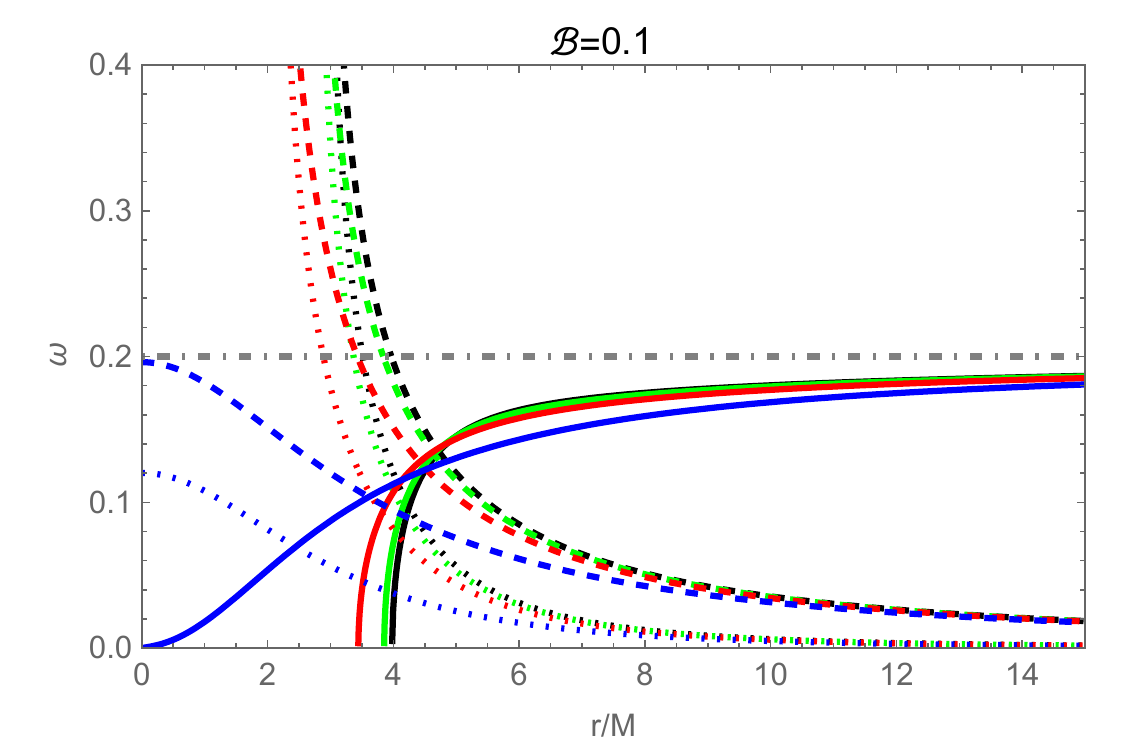}
\includegraphics[width=5.82cm]{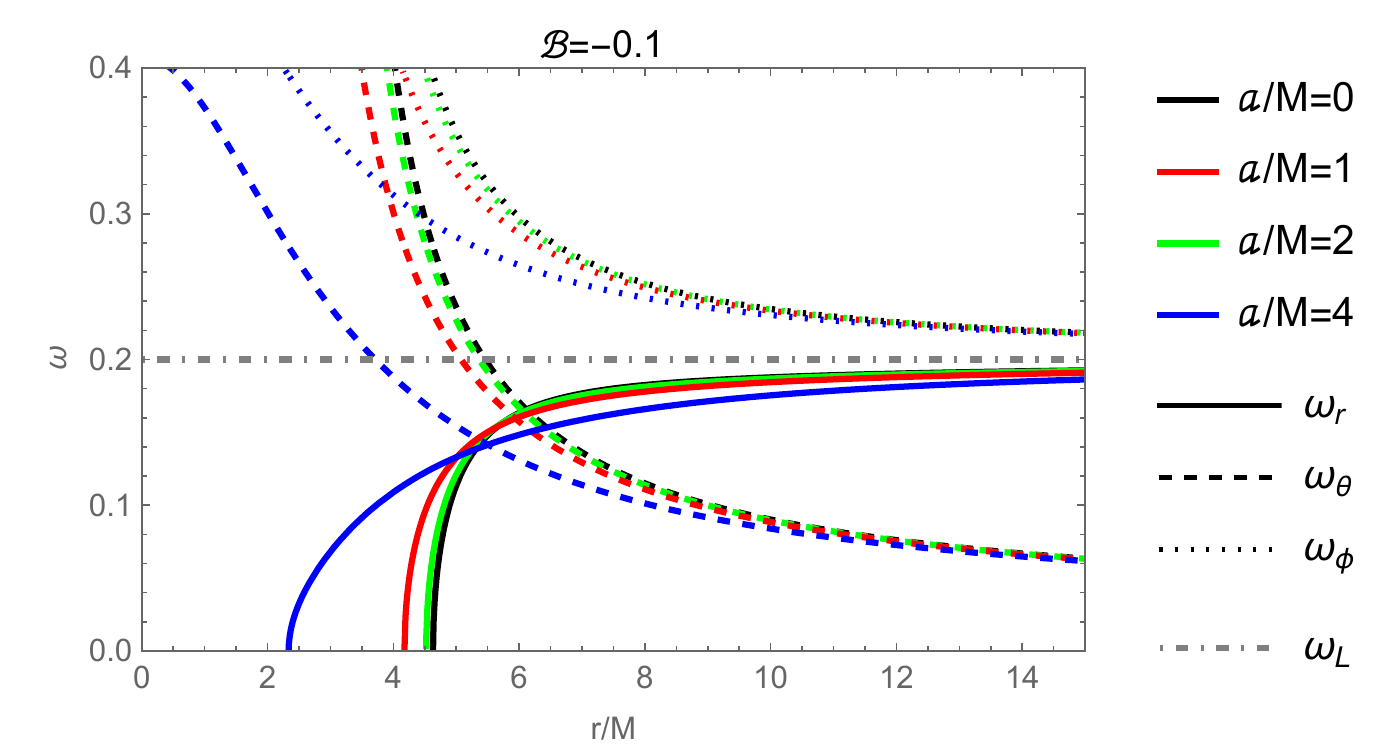}
  \caption{The variation curves of the angular frequency $\omega_r$ (solid line), $\omega_\theta$ (dashed line), $\omega_\phi$ (dotted line) of a test particle with respect to the radial coordinate $r$ in magnetized SV spacetime, under the influence of the regularization parameter $a$ and the magnetic parameter $\mathcal{B}$, along with the image of $\omega_L$ (dot-dash line).}
    \label{fig:2}
\end{figure}

In Figure \ref{fig:2}, we plot the curves showing the variation of the angular frequency $\omega_r$ (solid line), $\omega_\theta$ (dashed line) and $\omega_\phi$ (dotted line) of a test particle with respect to the radial coordinate $r$ in four types of magnetized SV spacetimes, under the influence of the regularization parameter $a$ and the magnetic parameter $\mathcal{B}$. Additionally, we also plot the image of $\omega_L$ (dot-dash line).

From Figure \ref{fig:2}, it is shown that if the value of $a$ is fixed, we have: $\omega_\theta=\omega_\phi$, the epicyclic radial, latitudinal angular frequencies, and the orbital frequency of the particle's motion around the central celestial body all approach to zero as $r$ increases,  when the SV spacetime is not immersed in a magnetic field $(\mathcal{B}=0)$; However, when the magnetic field is positive $(\mathcal{B}=0.1)$, we get: $\omega_\theta>\omega_\phi$, and as the radial coordinate increases, $\omega_r$ approach to $\omega_L$, while $\omega_\theta, \omega_\phi$ still tends to zero at infinity; For the case that the magnetic field is negative $(\mathcal{B}=-0.1)$, we have: $\omega_\theta<\omega_\phi$, in this case, the angular frequencies of the particle's motion $\omega_r, \omega_\phi $ tend to $\omega_L$ at infinity and $\omega_\theta$ tend to zero. Moreover, it can be observed that when the magnetic field direction and strength are fixed, the gradually increasing value of $a$ causes the values of $\omega_\theta, \omega_r$ and $\omega_\phi$ at the same radial coordinate position to decrease.

In the previous text, the derived $\omega_\theta, \omega_r, \omega_\phi$ represent the angular frequencies of the particle's local motion. To facilitate the correlation between theoretical values and observational data in the subsequent work, we need to study the angular frequency measured by a static observer at infinity. For this purpose, we perform the following transformation to equations (\ref{23})-(\ref{25}) with using the redshift factor:
\begin{equation}
\Omega_{r, \theta, \phi}=\frac{\omega_{r, \theta, \phi}}{f(r) \varepsilon(r)}, \label{27}
\end{equation}
here, $\varepsilon(r)$ specifically refers to the energy of the particle in circular orbit. Concretely, we gain
\begin{equation}
\Omega_\theta^2=\frac{1}{x^3}, \label{28}
\end{equation}
\begin{equation}
\Omega_\phi^2=\frac{1}{x^3\left\{1+2 \mathcal{B}\left[\mathcal{B}(-2+x) x^2+\sqrt{x^2\left[-3+x\left(1+\mathcal{B}^2(-2+x)^2 x\right)\right]}\right]\right\}}, \label{29}
\end{equation}
\begin{equation}
\begin{aligned}
&\begin{array}{r}
\Omega_r^2=\frac{1}{x^6\left\{-3+x-4 \mathcal{B}^2 x^2+2 \mathcal{B}^2 x^3-2 \mathcal{B} \sqrt{x} p\right\}} r^2\left[18+r^2-9 x+2 a^4 \mathcal{B}^2(-15+2 x)+a^2\left(1+\mathcal{B}^2(-48+\right.\right. \\
\left.\left.\left.68 x+r^2(-60+8 x)\right)\right)+2 \mathcal{B}\left(\mathcal{B} r^2\left(-24+34 x+r^2(-15+2 x)\right)-p(x-6)\right)\right]
\end{array}, \label{30}
\end{aligned}
\end{equation}
 with $p=\sqrt{x^2\left(-3+x+\mathcal{B}^2(-2+x)^2 x\right)}$. So far, the epicyclic motion and orbital motion of particle around a central celestial body have been considered as the primary sources of the QPOs phenomenon. Researchers are keen to utilize the angular frequencies of the particle' current cycle and orbital motion to construct relevant theoretical models, and combine these models with observed QPOs phenomena to study the QPO behavior of celestial bodies \cite{109}. To ensure that the physical quantities in the theoretical model are dimensionally consistent with the corresponding observational quantities, we define:
\begin{equation}
\nu=\frac{1}{2 \pi} \frac{c^3}{GM} \Omega_{r, \theta, \phi}. \label{31}
\end{equation}
In this context, $c$ is the speed of light, $G$ is the Newtonian gravitational constant, and $M$ is the mass of the celestial body.
\section{$\text{Discussion on resonance models and resonance positions in a magnetized SV spacetime}$}

The 3:2 ratio observed in the high $\left(\nu_u\right)$ and low $\left(\nu_l\right)$ peaks of the HFOPOs twin peaks in black holes and low-mass X-ray binaries (LMXBs) suggests that the HFQPO phenomenon may result from the resonance between various oscillation modes within the accretion disk \cite{132,133,135}. Therefore, considering the resonance of axisymmetric oscillation modes in the accretion disk, the epicyclic resonance (ER) model have been proposed \cite{60,61}. Geometrically, accretion disks are categorized into thin, thick, and toroidal accretion disks. For geometrically thin disks and elongated annular disks, the frequency of disk oscillation are related to the orbital and the epicyclic frequencies of circular geodesic motion \cite{136}. Generally, the ER model involves the parametric resonance model (PRM) and the forced resonance model (FRM), etc. Research indicates, objects with thin disks or those approximating Keplerian disks are more probable for PRM \cite{132,133,135}, and the PRM is governed by the Mathieu equation. At the same time, considering the possible dissipation, pressure effects, and other factors within the accretion disk \cite{137,139}, it is necessary to include forcing terms in the perturbation equation (\ref{20})  for test particle moving in the equatorial plane:
\begin{equation}
\delta \ddot{r}+\nu_{r}^2 \delta r=\nu_{r}^2\mathrm{~F}_{\mathrm{r}}(\delta r, \delta \theta, \delta \dot{r}, \delta \dot{\theta}), \delta \ddot{\theta}+\nu_\theta^2 \delta \theta=\nu_\theta^2\mathrm{~F}_\theta(\delta r, \delta \theta, \delta \dot{r}, \delta \dot{\theta}). \label{32}
\end{equation}
Here, $F_r$ and $F_\theta$ represent two undetermined functions corresponding to the coupling effects induced by the perturbation terms. In the PRM, it is assumed that $F_r=0, F_\theta=h \delta \theta \delta \mathrm{r}$, and $h$ is a constant \cite{140}. In this case, equation (\ref{32}) becomes:
\begin{equation}
\delta \ddot{r}+\nu_{r}^2 \delta r=0, \delta \ddot{\theta}+\nu_\theta^2\left[1+h \cos \left(\nu_{r} t\right)\right] \delta \theta=0, \label{33}
\end{equation}
and is excited when \cite{132,140,142}
\begin{equation}
\frac{\nu_r}{\nu_\theta}=\frac{2}{n},(n=1,2,3 \ldots) . \label{33}
\end{equation}
It is generally that frequencies corresponding to lower-order resonances are preferred, as they result in a larger amplitudes of the observed signal \cite{143}.

In a more realistic flow, however, the pressure, viscous, or magnetic stresses within the accretion flow can lead to the appearance of non-zero forced terms \cite{108,137}. Based on this, the FRM has been developed. In this model, the perturbation equation with non-zero forced terms can be written as:
\begin{equation}
\delta \ddot{\theta}+\nu_\theta^2 \delta \theta=-\nu_\theta^2 \delta r \delta \theta+\mathrm{F}_\theta(\delta \theta). \label{35}
\end{equation}
Where $\delta r=B \cos \left(\nu_r t\right), F_\theta$ denotes the nonlinear term related to $\delta \theta$. The forced resonance is activated when the epicyclic frequency ratio satisfies the following relationship:
\begin{equation}
\frac{\nu_\theta}{\nu_r}=\frac{p}{q}. \label{36}
\end{equation}

In the above equation, both $p$ and $q$ are small natural numbers. In the resonant solutions, the existence of combinational (beat) frequencies are permitted by the forced nonlinear resonance \cite{136}. Next, we assume that these compact objects exhibiting the observed HFQPOs phenomenon can be described by magnetized SV spacetime, and discuss how the aforementioned resonance models explain the 3:2 twin peak frequency ratio observed in the $X$-ray flux of these compact objects.

For SV spacetime not immersed in a magnetic field,  it always holds that $\nu_\phi=\nu_\theta>\nu_r$ when $r \geq r_{I S C O}$. Therefore, in the PRM, the minimum value of the resonance parameter is $n=3$ $\left(\nu_\theta: \nu_r=3: 2 \;\left(E R_0\right)\right)$; In FRM, it is required that $p / q$ is always greater than 1 (e.g. the lower order resonance $p: q= (3: 1\;(E R_1)$, and $2: 1\;(E R_3))$. In addition, considering the existence of other combined frequencies, there are $p: q= 5: 2\; \left(E R_2\right), \;5: 1 \;\left(E R_4\right), \;5: 3\; \left(E R_5\right)$, and the amplitude of oscillations is decreasing with increasing $p, q$.

For the case $\mathcal{B} \neq 0$, it still holds that $v_\theta>v_r$  near $r=r_{I S C O}$. However, as the radial coordinate increases,  both cases of $\nu_\theta=\nu_r$ and $\nu_\theta<\nu_r$ could occur, indicating that the presence of an external magnetic field could lead to the excitation of lower-order resonance parameters $\mathrm{n}=1 \left(\nu_\theta: \nu_r=\right.$ $\left.1: 2 \;\left(E R_6\right)\right)$, and $\mathrm{n}=2 \left(\nu_\theta: \nu_r=1: 1 \;\left(E R_7\right)\right)$. And in the FRM, this similarly implies the emergence of more frequency ratio scenarios (e.g., $p: q=1: 2\;\left(E R_8\right),\; p: q=2: 3\;\left(E R_9\right)$, etc.).

In order to achieve the observed frequency ratio $\nu_u: \nu_l=3: 2$, we posit that the combination of epicyclic frequencies and beat frequencies generates variants of ER model \cite{136}. In addition to the five common variants of the ER model: $E R_1-E R_5$ \cite{60,61}, we also considered other potential variants of the $E R$: $E R_6-E R_9$ that may exist in magnetized SV spacetime. For the considered models in this paper, one can see Table \ref{table1}.
\begin{table}[ht]
\begin{center}
\begin{tabular}{|c|c|c|c|c|c|c|c|c|c|c|}
\hline     & $E R_0$ & $E R_1$ & $E R_2$ & $E R_3$ & $E R_4$ & $E R_5$ & $E R_6$ & $E R_7$ & $E R_8$ & $E R_9$ \\
\hline$\nu_u$ & $\nu_\theta$ & $\nu_\theta$ & $\nu_\theta-\nu_r$ & $\nu_\theta+\nu_r$ & $\nu_\theta+\nu_r$ & $\nu_r$ & $3 \nu_\theta$ & $3 \nu_\theta$ & $\nu_\theta+\nu_r$ & $\nu_r$ \\
\hline$\nu_l$ & $\nu_r$ & $\nu_\theta-\nu_r$ & $\nu_r$ & $\nu_\theta$ & $\nu_\theta-\nu_r$ & $\nu_\theta-\nu_r$ & $\nu_r$ & $2 \nu_r$ & $\nu_r$ & $\nu_\theta$ \\
\hline
\end{tabular}
\end{center}
\caption{\label{table1}
 ER model and its variants considered in this paper.}
\end{table}

\begin{figure}[ht]
\includegraphics[width=4.99cm]{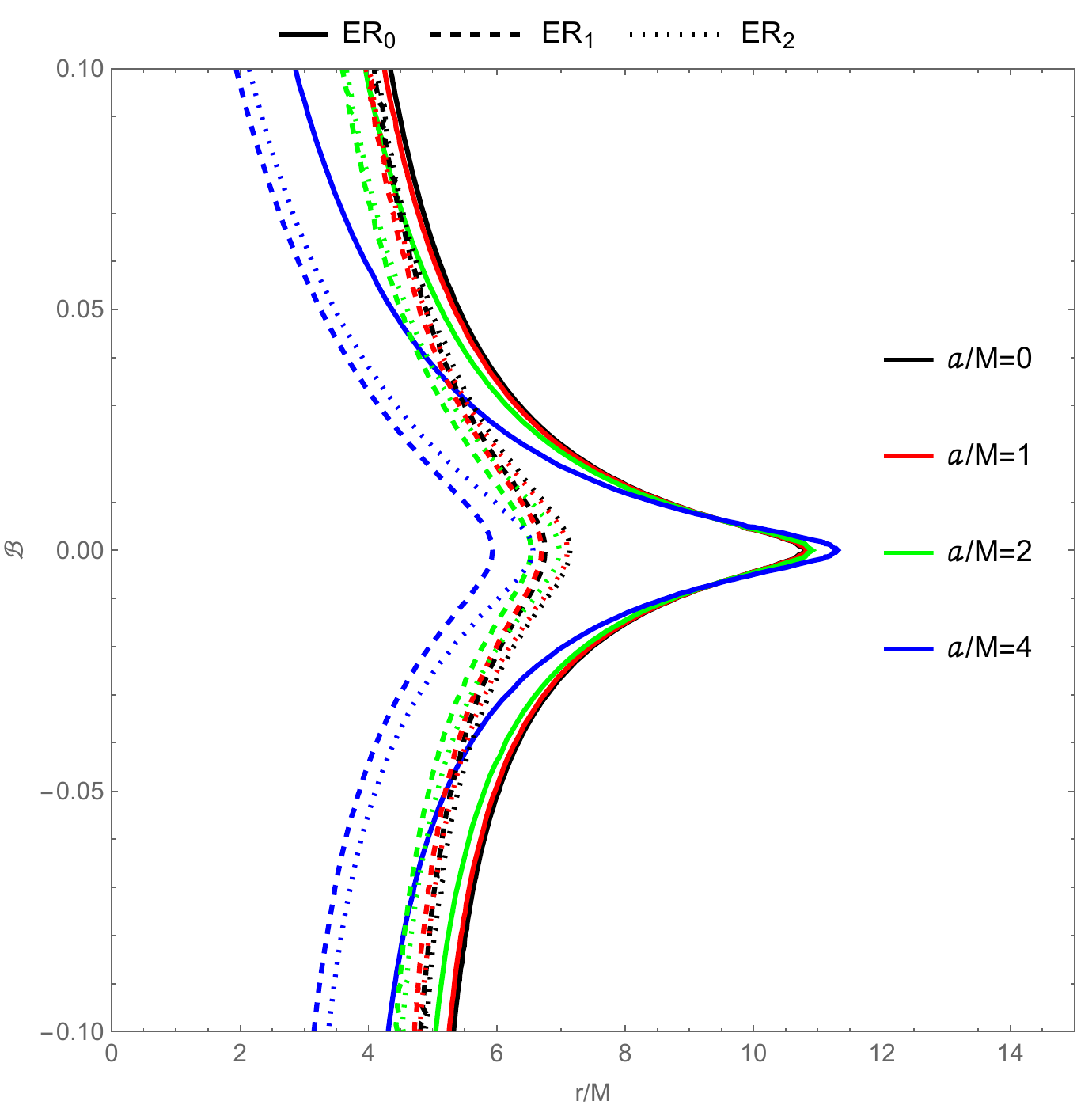}
\includegraphics[width=4.99cm]{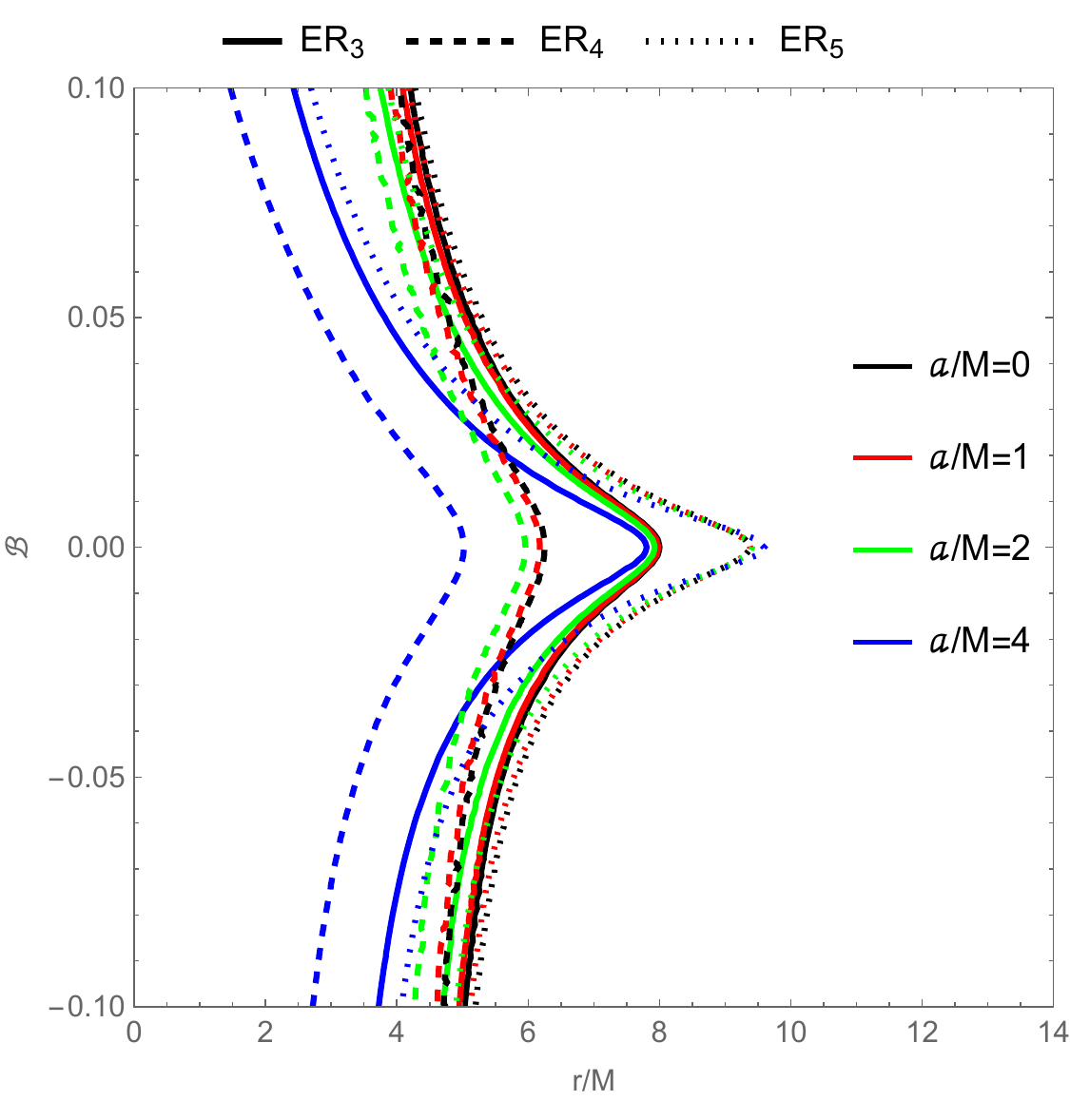}
\includegraphics[width=4.99cm]{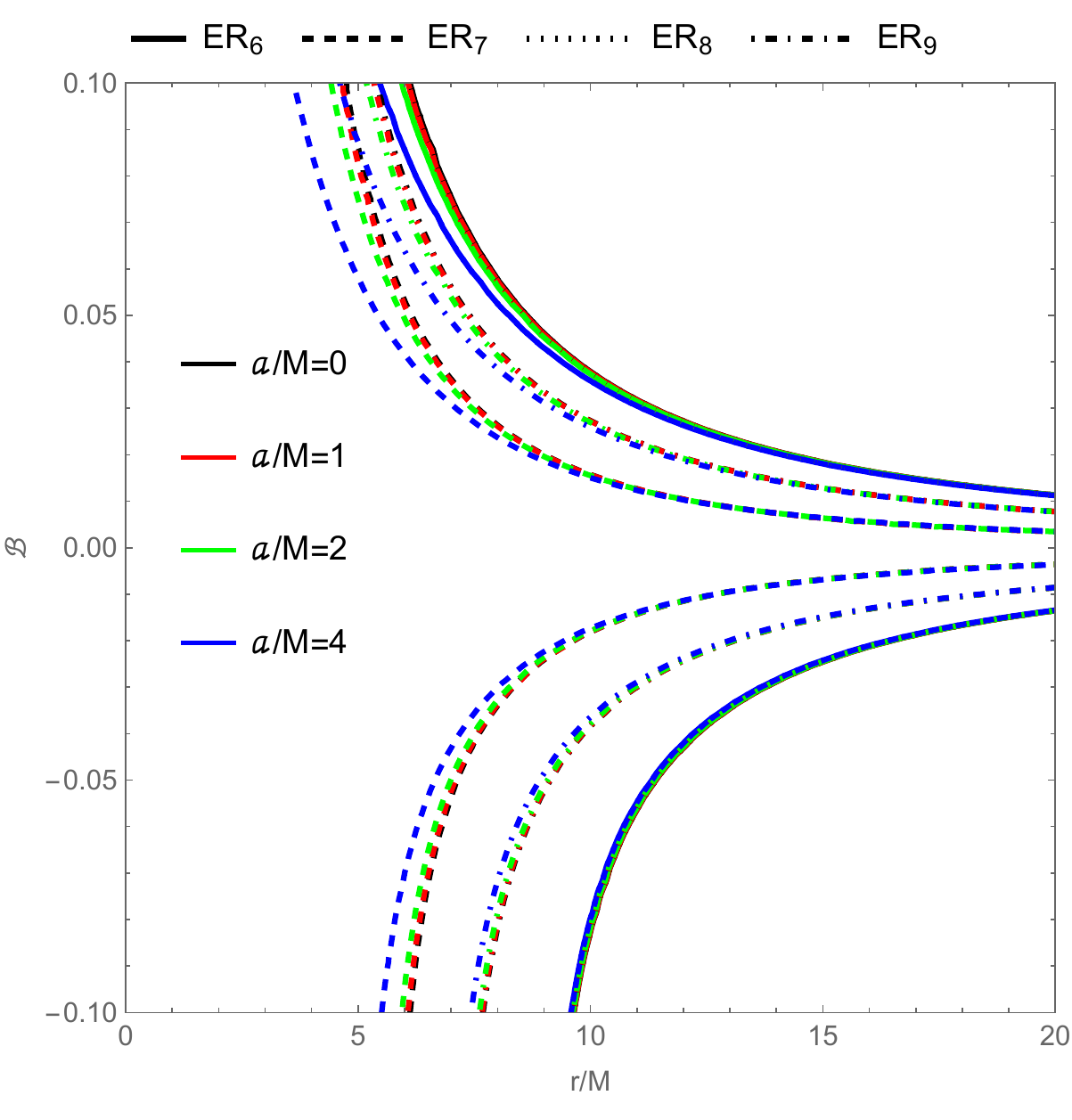}
   \caption{The variation of the $3: 2$ resonance position with the magnetic field parameter in four types of magnetized SV spacetimes under the resonance models $E R_0-E R_9$.}
    \label{fig:3}
\end{figure}

 Figure \ref{fig:3} shows the variation of the resonance positions with the magnetic field $\mathcal{B}$ in four types of magnetized SV spacetimes, considering the various resonance models listed in Table \ref{table1}. From Figure \ref{fig:3}, it can be observed that in the same type of magnetized SV spacetime, as the magnetic field $|\mathcal{B}|$ increases, the resonance position moves closer to the radial coordinate center. Furthermore, when the same strength $|\mathcal{B}|$ is set, the resonance radius for a magnetic field in the positive direction is smaller than that for a magnetic field in the negative direction, which is similar to the results in the ISCO.
\section{$\text{Constraints on the parameters of magnetized SV spacetime using HFQPOs observational data}$}

\subsection{{Data fitting and results}}
In recent years, it has been an important issue in relativistic astrophysics that the theory of gravity are tested using astronomical data. The used data include the HFQPOs observed in microquasars \cite{43,44,45,46,47,48,49,50}, the black hole shadow of Sgr A* observed by telescopes \cite{2,3}, and the hotspot data of three flares observed by gravitational instruments \cite{144}, among others. Of these observations, microquasars are considered as candidates for (stellar-mass) black holes. However, in practical situations, it is impossible to obtain precise information about these black holes. Therefore, we can test gravitational theories (black hole solutions) and the possible magnetic field configurations through observational data of HFQPOs \cite{145}.

In this section, we will use HFQPOs observational data from microquasars to constrain the regularization parameter $a/M$ and the magnetic parameter $\mathcal{B}$ in magnetized SV spacetime. We consider three sets of HFQPOs observational data from microquasars (see Table \ref{table2} for details) \cite{109,110,132}, which are labeled as: GRO 1655-40, XTE 1550-564, and GRS 1915+105. The specific data include the high and low frequencies of the HFQPOs twin peaks. In addition,the masses of the three microquasar sources $M / M_{\odot}$ were determined in \cite{147,148,149} via the optical/NIR photometry, where $M_{\odot}$ is the mass of the sun. We notice that, the epicyclic frequency models based on harmonic geodesic motion discussed in recent literature are incapable of simultaneously explain the HFQPOs in all three microquasars, assuming that their central attractor is a black hole \cite{76,110,146}. We expect that the presence of an external uniform magnetic field could improve this situation. Additionally, we teste the resonance models mentioned in Table \ref{table1}, aiming to find the HFQPO models that best matches with the observational data. This analysis will help us explore the possible physical mechanisms responsible for generating HFQPOs.
\begin{table}[ht]
\begin{center}
\begin{tabular}{|c|c|c|c|}
\hline     & GRO $1655-40\left(k_1\right)$ & XTE 1550-564 $\left(k_2\right)$ & GRS $1915+105\left(k_3\right)$ \\
\hline$\nu_u[\mathrm{~Hz}]$ & $441 \pm 2$ & $276 \pm 3$ & $168 \pm 3$ \\
\hline$\nu_l[\mathrm{~Hz}]$ & $298 \pm 4$ & $184 \pm 5$ & $113 \pm 5$ \\
\hline$M / M_{\odot}$ & $5.4 \pm 0.3$ & $9.1 \pm 0.61$ & $12.4_{-1.8}^{+2.0}$ \\
\hline
\end{tabular}
\end{center}
\caption{\label{table2}
 The observed HFQPOs data from three microquasars \cite{109,110,132}.}
\end{table}

\begin{figure}[ht]
\includegraphics[width=4.8cm]{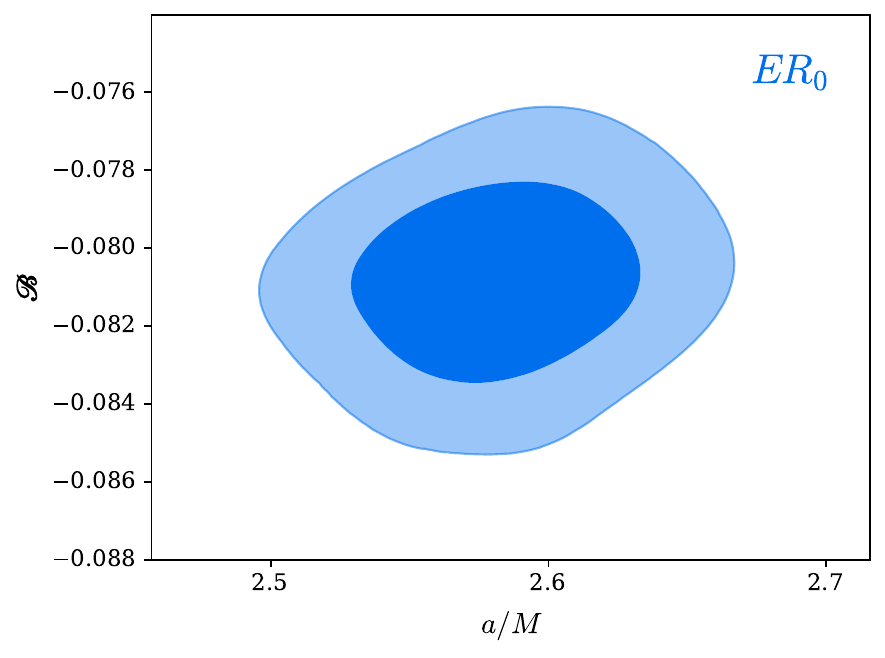}
\includegraphics[width=4.8cm]{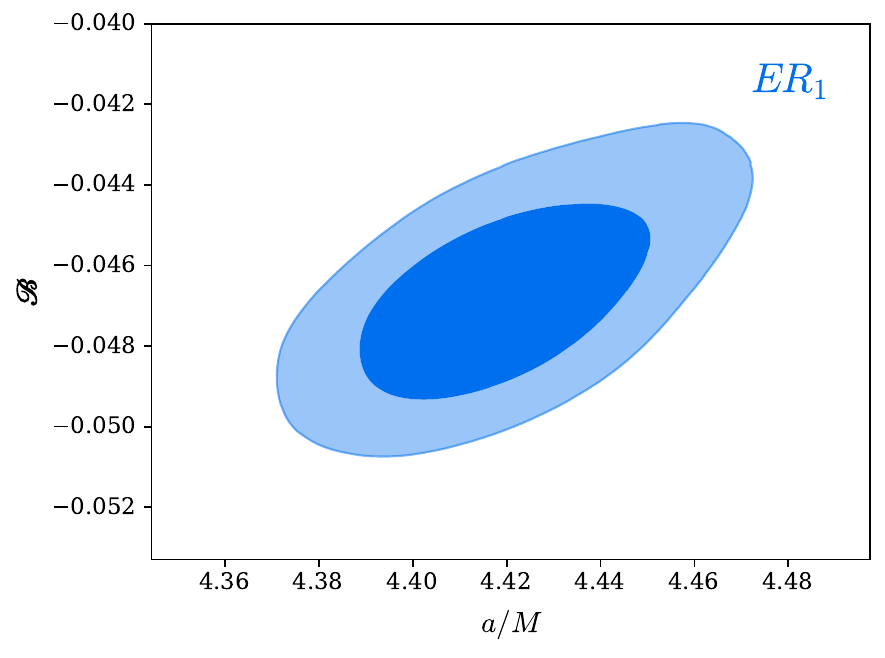}
\includegraphics[width=4.8cm]{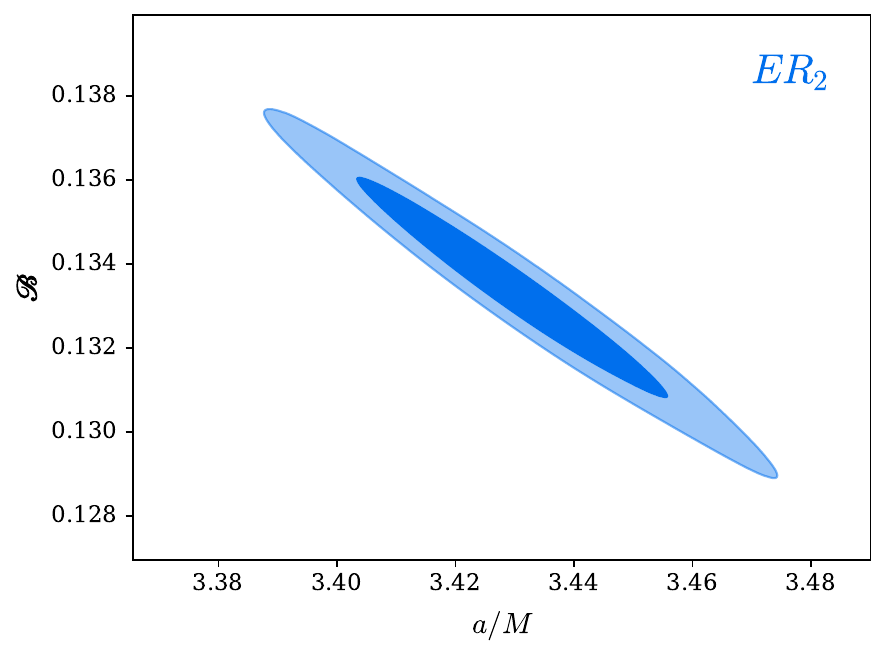}
\includegraphics[width=4.8cm]{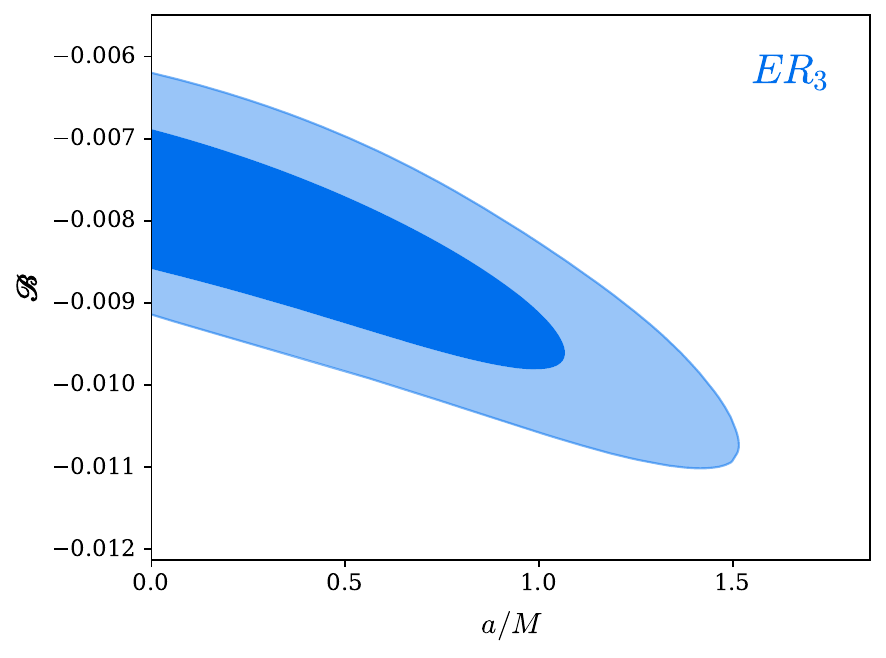}
\includegraphics[width=4.8cm]{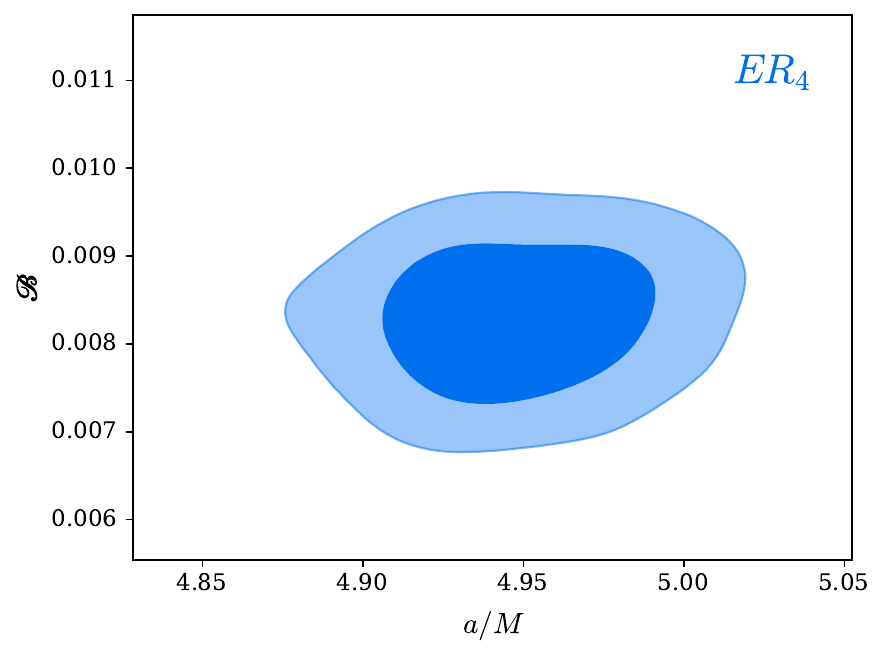}
\includegraphics[width=4.8cm]{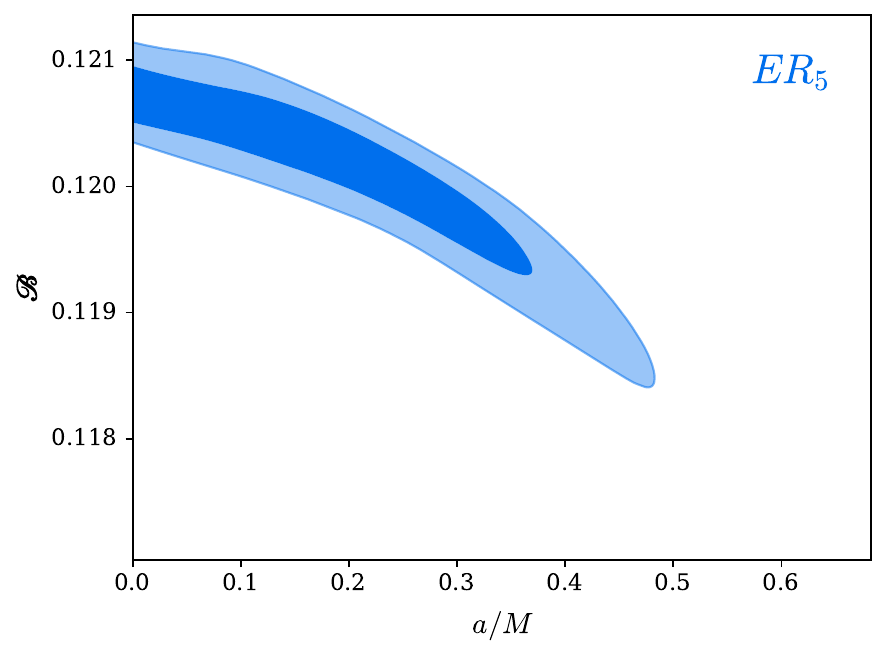}
\caption{The $68 \%$ and $95 \%$ confidence regions of the magnetic parameter $\mathcal{B}$ and the regularization parameter $a$ under $E R_0-E R_5$ resonance models.}
    \label{fig:4}
\end{figure}

\begin{figure}[ht]
\includegraphics[width=4.2cm]{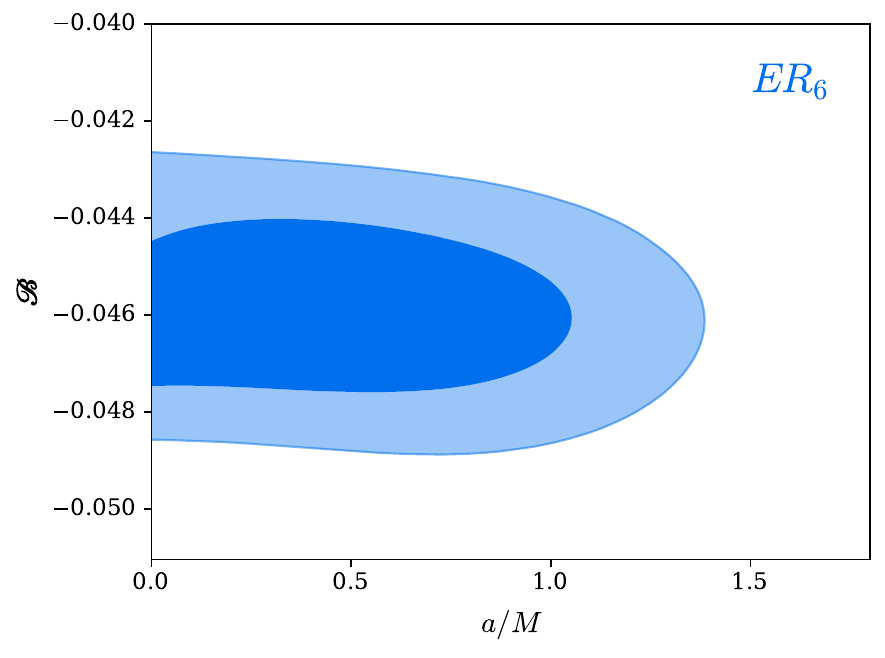}
\includegraphics[width=4.2cm]{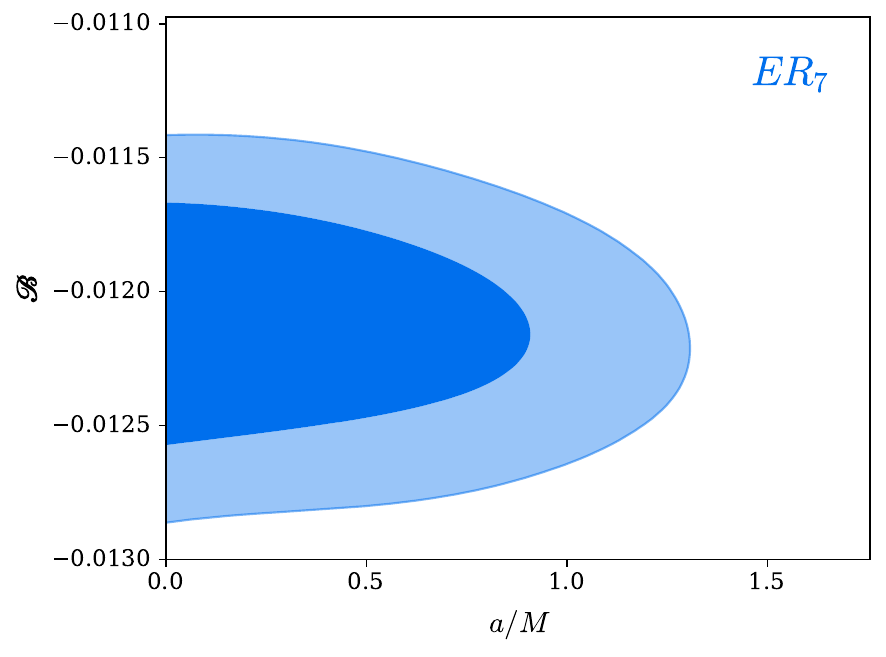}
\includegraphics[width=4.2cm]{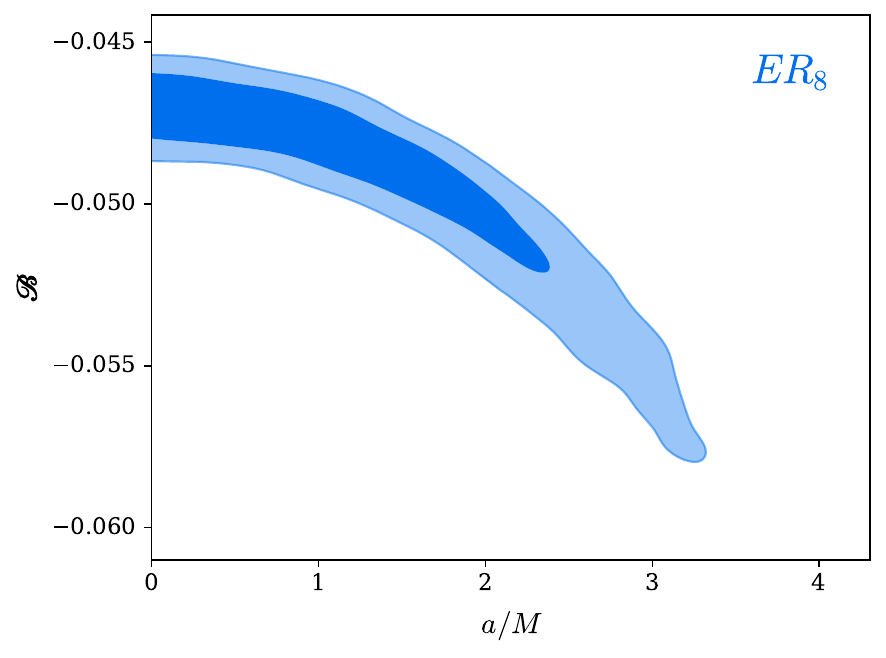}
\includegraphics[width=4.2cm]{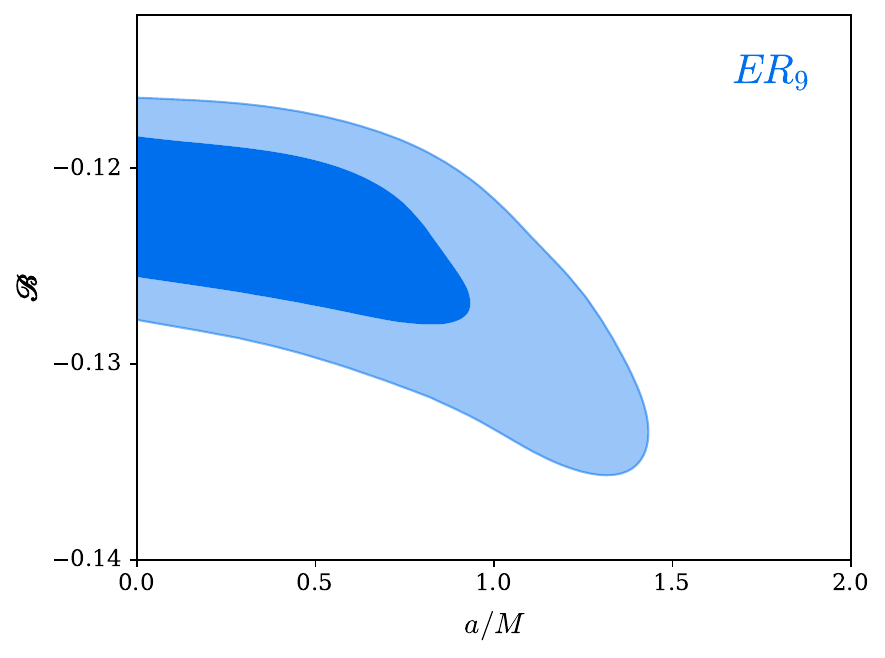}
\caption{The $68 \%$ and $95 \%$ confidence regions of the magnetic parameter $\mathcal{B}$ and the regularization parameter $a$ under $E R_6-E R_9$ resonance models.}
\label{fig:5}
\end{figure}

 We consider that the three sets of HFQPOs are generated at the different position of circular orbits with radii $r_1^*$,$r_2^*$,$r_3^*$ $\left(r_{1,2,3}^*=r_{1,2,3} / M\right)$, respectively. From equations (\ref{28}) to (\ref{31}), it can be seen that there are five free parameters for the theoretical model in total: the regularization parameter $a/M$, the magnetic field parameter $\mathcal{B}$, and the position of the 3:2 resonance $r_k^*(k=1,2,3)$. To determine the values of model parameters, with using the above observational data, we perform a $\chi^2$ analysis:
\begin{equation}
\chi^2=\sum_{k=1}^3 \frac{\left\{\nu_{u, k}-\nu_u\left(a/M, \mathcal{B}, r_k^*\right)\right\}^2}{\sigma_{\nu_{u, k}}^2}+\sum_{k=1}^3 \frac{\left\{\nu_{l, k}-\nu_l\left(a/M, \mathcal{B}, r_k^*\right)\right\}^2}{\sigma_{\nu_{l, k}}^2} . \label{37}
\end{equation}
 According to the constraint results from the HFQPO data, we summarize the best-fit values of the magnetic parameter $\mathcal{B}$ and the regularization parameter $a/M$ within the $68 \%$ confidence interval, as well as the minimum $\chi^2$ value, in Table \ref{table3}. Furthermore, we show the best-fit values of the circular orbit radii associated with the three sets of QPO data, i.e., the resonance positions, which are listed in Table \ref{table4}. In Figure \ref{fig:4} and \ref{fig:5}, we plot the $68 \%$ and $95 \%$ confidence levels (CL) for magnetic parameter $\mathcal{B}$ and regularization parameter $a/M$ with different resonance models.

\begin{table}[ht]
\begin{center}
\begin{tabular}{|c|c|c|c|c|c|}
\hline
\multirow{3}{*}{$E R_0$}        & $a / M$ & $2.589 \pm 0.036$ &                        \multirow{3}{*}{$E R_1$} & $a / M$ & $4.420_{-0.021}^{+0.018}$ \\ \cline{2-3}\cline{5-6}
                                & $\mathcal{B}$ & $-0.0809 \pm 0.0022$ &                           & $\mathcal{B}$ & $-0.0466 \pm 0.0016$ \\ \cline{2-3}\cline{5-6}
                                & $\chi_{\text {min }}^2$ & 52.584 &                   & $\chi_{\text {min }}^2$ & 22.711 \\ \cline{2-3}\cline{5-6}

\hline \multirow{3}{*}{$E R_2$} & $a / M$ & $3.425 \pm 0.015$ & \multirow{3}{*}{$E R_3$} & $a / M$ & $<0.736$ \\
\cline{2-3}\cline{5-6} & $\mathcal{B}$ & $0.1339 \pm 0.0016$ & & $\mathcal{B}$ & $-0.0086_{-0.00069}^{+0.00095}$ \\
\cline{2-3}\cline{5-6} & $\chi_{\text {min }}^2$ & 46.517 & & $\chi_{\text {min }}^2$ & 5.84 \\

\hline \multirow{3}{*}{$E R_4$} & $a / M$ & $4.956_{-0.027}^{+0.024}$ & \multirow{3}{*}{$E R_5$} & $a / M$ & $<0.225$ \\
\cline{2-3}\cline{5-6} & $\mathcal{B}$ & $0.00843 \pm 0.00062$ & & $\mathcal{B}$ & $-0.12007_{-0.00033}^{+0.0008}$ \\
\cline{2-3}\cline{5-6} & $\chi_{\text {min }}^2$ & 55.796 & & $\chi_{\text {min }}^2$ & 74.771 \\

\hline \multirow{3}{*}{$E R_6$} & $a / M$ & $0.6_{-0.59}^{+0.16}$ & \multirow{3}{*}{$E R_7$} & $a / M$ & $<0.602$ \\
\cline{2-3}\cline{5-6} & $\mathcal{B}$ & $-0.0461 \pm 0.0013$ & & $\mathcal{B}$ & $-0.01211 \pm 0.00032$ \\
\cline{2-3}\cline{5-6} & $\chi_{\text {min }}^2$ & 7.658 & & $\chi_{\text {min }}^2$ & 7.161 \\

\hline \multirow{3}{*}{$E R_8$} & $a / M$ & $<1.77$ & \multirow{3}{*}{$E R_9$} & $a / M$ & $<0.649$ \\
\cline{2-3}\cline{5-6} & $\mathcal{B}$ & $-0.0493_{-0.000072}^{+0.0031}$ & & $\mathcal{B}$ & $-0.1240_{-0.0024}^{+0.0039}$ \\
\cline{2-3}\cline{5-6} & $\chi_{\text {min }}^2$ & 2.623 & & $\chi_{\text {min }}^2$ & 61.827 \\
\hline
\end{tabular}
\end{center}
\caption{\label{table3}
 The best-fit values of the magnetic parameter $\mathcal{B}$ and the regularization parameter $a/M$ within the 68\% confidence level, as well as the minimum $\chi^2$ value, for different resonance models.}
\end{table}

\begin{table}[ht]
\begin{center}
\begin{tabular}{|c|c|c|c|c|c|c|c|c|c|c|}
\hline & $E R_0$ & $E R_1$ & $E R_2$ & $E R_3$ & $E R_4$ & $E R_5$ & $E R_6$ & $E R_7$ & $E R_8$ & $E R_9$ \\
\hline$r_1^*$ & 5.083 & 3.600 & 2.150 & 7.400 & 4.071 & 4.044 & 11.842 & 11.853 & 12.104 & 7.413 \\
\hline$r_2^*$ & 4.938 & 3.421 & 2.110 & 7.022 & 3.904 & 4.058 & 11.440 & 11.455 & 11.188 & 7.224 \\
\hline$r_3^*$ & 5.477 & 4.104 & 2.244 & 8.462 & 4.992 & 4.016 & 13.027 & 13.026 & 14.718 & 9.255 \\
\hline
\end{tabular}
\end{center}
\caption{\label{table4}
 The best-fit values of the circular orbit radiis associated with the three sets of QPOs under different resonance models, i.e., the orbital radii that produce the three sets of HFQPOs (resonance positions).}
\end{table}

\subsection{{Model selection}}

\subsubsection{{Analysis from Akaike Information Criterion and Bayesian Information Criterion}}
In this part, we utilize the obtained $\chi_{\text {min }}^2$values to calculate the AIC for evaluating the fit of different resonance models with the three sets of HFQPOs data, in order to determine which models the observational data favor. In cosmology, the AIC was first introduced by Liddle \cite{150}, and then was generalized to other studies \cite{151,152}. It is defined as \cite{153}
\begin{equation}
\mathrm{AIC}=-2 \ln \mathcal{L}(\hat{\theta} \mid \text { data })_{\max }+2 K, \label{38}
\end{equation}
where $K$ is the number of estimable parameters $\hat{\theta}$ in the model and $\mathcal{L}_{\text {max }}$ is the highest likelihood in the model with the best fit parameters $\hat{\theta}$. The term $-2 \ln \mathcal{L}(\hat{\theta} \mid \text { data })_{\max }$ in Eq.(\ref{38}) is called $\chi^2$ and it measures the quality of model fit, while the term $2 K$ in Eq.(\ref{38}) interprets model complexity \cite{151,152,154,157,158}.

The AIC value of a single model is not meaningful in itself, only the relative values among different models hold practical significance. Therefore, the model with the minimized value of AIC is regarded as the optimal model, denoted as $\mathrm{AIC}_{\min }=\min \left\{\mathrm{AIC}_i, i=1, \ldots, N\right\}$, and $\mathrm{AIC}_i$ is a set of alternative candidate models. By calculating the likelihood of the model $\mathcal{L}\left(M_i \mid\right.$ data $) \propto \exp \left(-\Delta_i / 2\right)$, one can ascertain the relative evidential strength of each model, where $\Delta_i=\mathrm{AIC}_i-\mathrm{AIC}_{\text {min }}$ over the whole range of alternative models. The criteria for evaluating model selection are as follows: when $0 \leqslant \Delta_i < 2$, model $i$ receives nearly the same level of support from the data as the best model; for $2 \leqslant \Delta_i < 4$, there is a modest level of evidence favoring model $i$,  for $4 \leqslant \Delta_i < 7$, model $i$ has significantly less support; and when $\Delta_i>10$, model $i$ is essentially irrelevant.

In terms of the calculation results, the best model is $E R_8\left(\nu_u=\nu_\theta+\nu_r, \nu_l=\nu_r\right)$, with an AIC value of 12.623. Using it as a reference, we can compare the models in Table 1 by employing the AIC difference $\triangle_i$. Considering that all the HFQPO models in Table 1 have the same number of model parameters, the AIC difference $\Delta_i$ can also be calculated by the $\Delta \chi_{\min }^2$ values. It's important to recognize that AIC model selection offers quantitative insights into the "strength of evidence," rather than simply identifying a single best model. We present the HFQPO models within different $\triangle_i$ intervals in Table \ref{table5}.
\begin{table}[ht]
\begin{center}
\begin{tabular}{|c|c|c|c|c|}
\hline      & $0 \leqslant \Delta_i < 2$ & $2\leqslant\Delta_i < 4$ & $4\leqslant \Delta_i < 7$ & $\Delta_i>10$ \\
\hline HFQPO models & $E R_8$ & $E R_3$ & $E R_6, E R_7$ & \begin{tabular}{l}
$E R_0 , E R_1 , E R_2 , E R_4, $ \\
$E R_5 , E R_9$
\end{tabular} \\
\hline
\end{tabular}
\end{center}
\caption{\label{table5}
 Relative to the best model $ER_8$, the $\Delta_i$ intervals for AIC for other HFQPO models.}
\end{table}

From Table \ref{table5}, it can be seen that the differences in support for different theoretical models by the three sets of HFQPOs data are significant. Among them, $E R_8$ model is the best model and has received strong support, $ER_3$ model has received moderate evidence of support, whereas $ER_6$, $ER_7$ models have considerably less support. Furthermore, there is no substantial evidence in favor of the other resonance models mentioned in Table \ref{table1}. Through Tables \ref{table3} and \ref{table5}, we find that for the models more supported by the observational data ($0 \leqslant \Delta_i < 4$), the common interval for the regularization parameter $0\leq a<0.736$ ($68\%$ CL). This suggests that the HFQPOs observational data support the magnetized black bounce spacetime as a regular black hole, and the small values of $a$ might also reflect certain quantum effects.

In addition to the AIC, the BIC proposed by Schwarz, further takes into account the effect of the number of observational data points on the penalty for model complexity. The BIC is defined as \cite{BIC}
\begin{equation}
\mathrm{BIC}=-2 \ln \mathcal{L}(\hat{\theta} \mid \text { data })_{\max }+K\ln N,  \label{39}
\end{equation}
where $N$ is the number of observational data points used in the fitting procedure. Compared with the AIC, the complexity penalty in the BIC depends on both the number of model parameters and the sample size, and therefore the BIC can serve as a complementary criterion to further assess the consistency and robustness of the model-selection results. It should be noted that the different QPO models considered in this work have the same number of free parameters and are fitted to the same observational dataset. Therefore, the values $\Delta_i =\mathrm{BIC}_i-\mathrm{BIC}_{\text {min }}$ derived from the BIC are identical to those obtained from the AIC. However, the criterion for interpreting the BIC slightly differs from that commonly adopted for the AIC. The BIC model selection criterion are as follows: $\Delta_i<2$ (e.g. $ER_8$) indicates weak or negligible evidence against the model, $2\leq\Delta_i<6$ (e.g. $ER_3, ER_6, ER_7$) suggests positive evidence, $6\leq\Delta_i<10$ shows strong evidence  against the model, and $\Delta_i\geq10$ (e.g. $ER_0, ER_1, ER_2, ER_4, ER_5, ER_9$) demonstrates very strong evidence against the model relative to the best-fitting model. Clearly, the BIC-based evaluation of the QPO models considered in this work yields results that are broadly consistent with those obtained using the AIC.

\subsubsection{{Bayesian analysis}}

To further compare the agreement between different resonance models and the observational data, we perform a Bayesian model comparison in this section. For a specified spacetime model $\bar{M}$, the inference of the parameter set $\Theta$ is based on the posterior distribution conditioned on the data $D$. According to Bayes' theorem, this distribution can be expressed as \cite{Bys-1, Bys-2}
\begin{equation}
P(\Theta \mid D, \bar{M})=\frac{P(D \mid \Theta, \bar{M}) \pi(\Theta \mid \bar{M})}{P(D \mid \bar{M})},
\end{equation}
where $P(D\mid\Theta,\bar{M})$ is the likelihood associated with the parameter values $\Theta$, while $\pi(\Theta\mid\bar{M})$ denotes their prior distribution under the model $\bar{M}$. The quantity appearing in the denominator is the marginal likelihood of the model, which is obtained by integrating over the full parameter space,
\begin{equation}
P(D \mid \bar{M})=\int P(D \mid \Theta, \bar{M}) \pi(\Theta\mid\bar{M}) d \Theta.
\end{equation}

The probability assigned to the model itself after taking the data into account is then related to its evidence through
\begin{equation}
P(\bar{M} \mid D) \propto P(D \mid \bar{M}) \pi(\bar{M}) \propto Z_{\bar{M}} \pi_{\bar{M}},
\end{equation}
with $Z_{\bar{M}}=P(D\mid\bar{M})$ representing the Bayesian evidence and $\pi_{\bar{M}}=\pi(\bar{M})$ denoting the prior probability of the model. Therefore, for two competing spacetime models, $\bar{M}_1$ and $\bar{M}_2$, their relative support from the data can be quantified through
\begin{equation}
R=\frac{Z_{\bar{M}_1} \pi_{\bar{M}_1}}{Z_{\bar{M}_2} \pi_{\bar{M}_2}}.
\end{equation}
If the two models are assigned identical prior probabilities, the above ratio is reduced to the usual Bayes factor. In this case, $R>1$ means that the observational data provide greater support for $\bar{M}_1$, whereas $R<1$ favors $\bar{M}_2$ \cite{Bys-1}.

The degree to which one model is preferred over the other can be further interpreted using the empirical Jeffreys scale \cite{Bys-3,Bys-4,Bys-5}. This classification is based on the magnitude of $|\ln R|$. When $|\ln R|<1$, no meaningful preference can be established between the two models. The interval $1\leq|\ln R|<2.5$ corresponds to weak support, whereas $2.5\leq|\ln R|<5$ indicates moderate support. A value satisfying $5\leq|\ln R|<10$ is commonly regarded as strong evidence, while $|\ln R|\geq10$ signifies decisive support for one model relative to the other.

In Table \ref{table6}, we present the marginal likelihoods and the corresponding log-evidence values for the $ER_0$--$ER_9$ models. Among them, $ER_8$ yields the largest Bayesian evidence and is therefore chosen as the reference model. We then evaluate its relative statistical preference over the remaining models in terms of the Bayes factor $R$ and the corresponding logarithmic quantity $\ln R$. The resulting values of $R$ and $\ln R$ are reported in \ref{table7}.

\begin{table}[ht]
\begin{center}
\begin{tabular}[t]{|l|l|l|l|l|l|l|l|l|l|l|}
\hline & $E R_0$ & $E R_1$ & $E R_2$ & $E R_3$ & $E R_4$ & $E R_5$ & $E R_6$ & $E R_7$ & $E R_8$ & $E R_9$ \\
\hline $Z$ & $2.96 * 10^{-12}$ & $9.15 * 10^{-6}$ & $5.90 * 10^{-11}$ & 0.053 & $5.84 * 10^{-13}$ & $5.38 * 10^{-17}$ & 0.020 & 0.024 & 0.237 & $3.180 * 10^{-14}$ \\
\hline $|\operatorname{Ln} Z|$ & -26.54 & -11.60 & -23.55 & -2.93 & -28.17 & -37.46 & -3.908 & -3.709 & -1.441 & -31.079 \\
\hline
\end{tabular}
\end{center}
\caption{\label{table6}
 The marginal likelihoods and the corresponding log-evidence values for the $ER_0$-$ER_9$ models.}
\end{table}

\begin{table}[ht]
\begin{center}
\begin{tabular}[t]{|l|l|l|l|l|l|l|l|l|l|}
\hline & $E R_8 / E R_0$ & $E R_8 / E R_1$ & $E R_8 / E R_2$ & $E R_8 / E R_3$ & $E R_8 / E R_4$ & $E R_8 / E R_5$ & $E R_8 / E R_6$ & $E R_8 / E R_7$ & $E R_8 / E R_9$ \\
\hline $R$ & $8.007 * 10^{10}$ & $2.590 * 10^4$ & $4.017 * 10^9$ & 4.472 & $4.058 * 10^{11}$ & $4.405 * 10^{15}$ & 11.850 & 9.875 & $7.453 * 10^{12}$ \\
\hline $|\operatorname{Ln} R|$ & 25.099 & 10.159 & 22.109 & 1.489 & 26.729 & 36.019 & 2.467 & 2.268 & 29.638 \\
\hline
\end{tabular}
\end{center}
\caption{\label{table7}
The values of the Bayes factors and $|\ln R|$ of different QPO models relative to the $ER_8$ model.}
\end{table}

Overall, $ER_8$ has the largest Bayesian evidence and is therefore the most favored model among those considered. The Bayesian comparison provides decisive evidence for $ER_8$ over $ER_0$, $ER_1$, $ER_2$, $ER_4$, $ER_5$, and $ER_9$, whereas weak evidence is found relative to $ER_3$, $ER_6$, and $ER_7$. Thus, these three models remain,  to some extent, statistically competitive with $ER_8$. Obviously, the Bayesian analysis yields model-selection results that are broadly consistent with those obtained using the AIC and BIC methods.

\subsection{{Discussions on magnetic field strength from observational results}}
In the preceding text, we utilize observational data of HFQPOs from microquasars to constrain the value of the magnetic parameter $\mathcal{B}$ under the assumption of a uniform magnetic field surrounding the SV spacetime. In actual scenarios, the real magnetic fields surrounding microquasars or supermassive black holes and their accretion disks are not entirely regular and uniform. In light of this, the Wald uniform magnetic field solution was introduced as an approximation capable of accurately describing the magnitude of the magnetic field. In the context of the QPO models, it is assumed that a uniform magnetic field is sufficient \cite{73,76}.

As can be observed from equation (\ref{12}), the magnetic parameters, in conjunction with the field strength, also incorporate the specific charge of the ionized test particle. This implies that, in order to accurately estimate the magnitude of the magnetic field, we must ascertain the type of matter within the accretion disc. It is well known that, due to the high temperatures of the accretion disk surrounding black holes, the astrophysical plasma is composed of ions and electrons. Among these, hydrogen is likely the most abundant ion in the disk \cite{159}. Therefore, in this article, we select electrons and protons as specific examples. In the dimensionless magnetic parameters, the physical parameter is referred to as $\mathcal{B}=|q| B G M /\left(2 m c^4\right)$, and the magnetic field strength in Gauss can be derived from equation (\ref{12}) \cite{60,159}:
\begin{equation}
B=\frac{2 m c^4 \mathcal{B}}{q G M} . \label{39}
\end{equation}
In Table \ref{table6}, we present the magnetic field strengths around the three microquasars in the HFQPO models where $0 \leqslant \Delta_i < 6$ for the Information-Criterions evaluation and $1 \leq |\operatorname{Ln} R|<2.5$ for the Bayesian analysis (i.e. for models supported by the data), considering the cases when the test particle is an electron or a proton. The quantities are given in GS units.
\begin{table}[ht]
\begin{center}
\begin{tabular}{|c|c|c|c|c|c|}
\hline \multicolumn{2}{|c|}{$B$} & $E R_8$ & $E R_3$ & $E R_6$ & $E R_7$ \\
\hline \multirow{2}{*}{ GRO 1655-40} & Electron & $2.105 * 10^{-4}$ & $3.671 * 10^{-5}$ & $1.968 * 10^{-4}$ & $5.170 * 10^{-5}$ \\
\cline { 2 - 6 } & Proton & 0.3864 & 0.06741 & 0.3614 & 0.09492 \\
\hline \multirow{2}{*}{ XTE 1550-564 } & Electron & $1.249 * 10^{-4}$ & $2.179 * 10^{-5}$ & $1.168 * 10^{-4}$ & $3.068 * 10^{-5}$ \\
\cline { 2 - 6 } & Proton & 0.2293 & 0.04000 & 0.2144 & 0.05632 \\
\hline \multirow{2}{*}{ GRS 1915+105 } & Electron & $9.165 * 10^{-5}$ & $1.599 * 10^{-5}$ & $5.571 * 10^{-5}$ & $2.251 * 10^{-5}$ \\
\cline { 2 - 6 } & Proton & 0.1683 & 0.02936 & 0.1574 & 0.04134 \\
\hline
\end{tabular}
\end{center}
\caption{\label{table8}
 The best fit values of magnetic field strength (GS) around the three microquasars for the HFQPO models  supported by the observational data, when the test particle is an electron or a proton.}
\end{table}

From Table \ref{table8}, it can be found that when the test particle is assumed to be an electron, the best fit values of magnetic field strength around the three microquasars approximately $10^{-5}\sim 10^{-4}$ GS, for the HFQPO models  supported by the observational data. This is consistent with the measured values of large-scale Galactic magnetic fields, and also aligns with the estimates of large-scale uniform magnetic fields around microquasars, assuming they are represented by Kerr spacetime, as reported in the literature \cite{76}. Additionally, when the test particle is assumed to be a proton, the estimated magnetic field strength in the $E R_8$ and $E R_6$ models are comparable to the magnetic field at the Earth's surface \cite{90}, while in the $E R_3, E R_7$ models, the estimated magnetic field strength is an order of magnitude smaller than the former. Finally, we propose that in real scenarios, with a fixed magnetic field strength $B$ (or with small fluctuations), the significant influence of the Lorentz force on charged particle could lead to substantial differences in the trajectories of particle with different charge-to-mass ratios moving around compact objects. For example, as discussed in this paper, under a fixed magnetic field strength $B$, protons may remain in circular orbits within the plane of the accretion disk, while electrons might escape from the accretion disk along magnetic field lines.

\subsection{{No magnetic field case}}
\begin{figure}[ht]
\includegraphics[width=4.8cm]{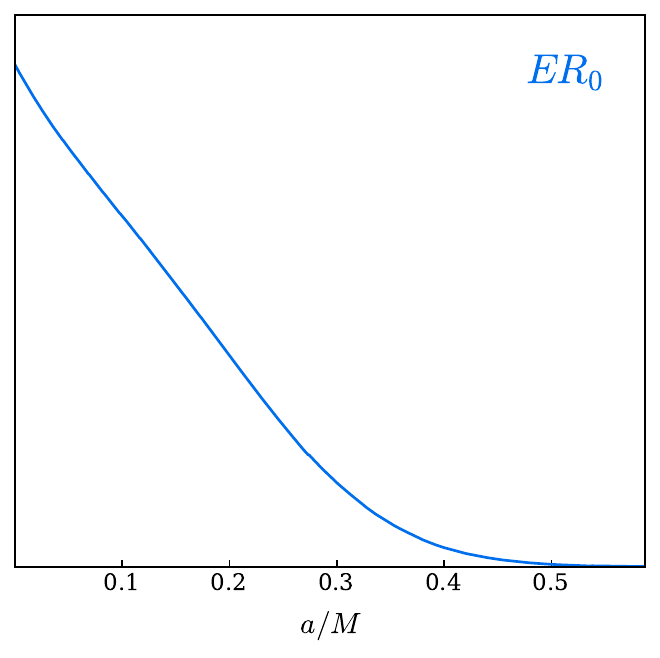}
\includegraphics[width=4.8cm]{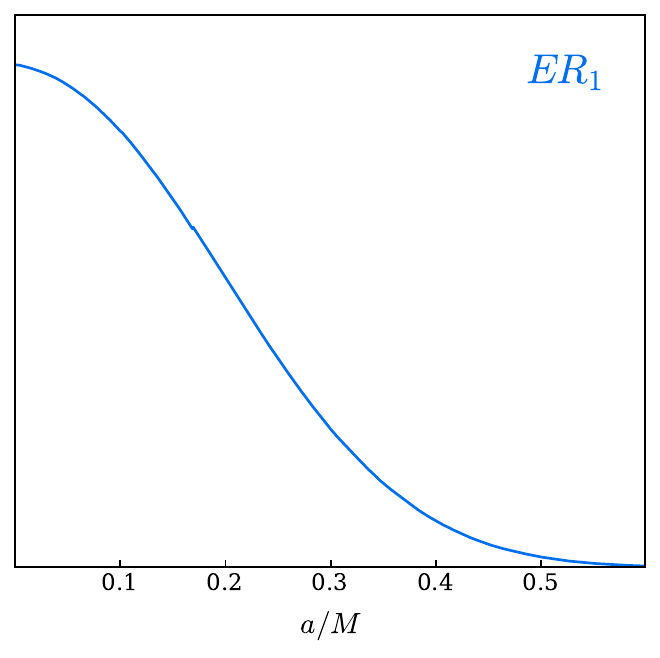}
\includegraphics[width=4.8cm]{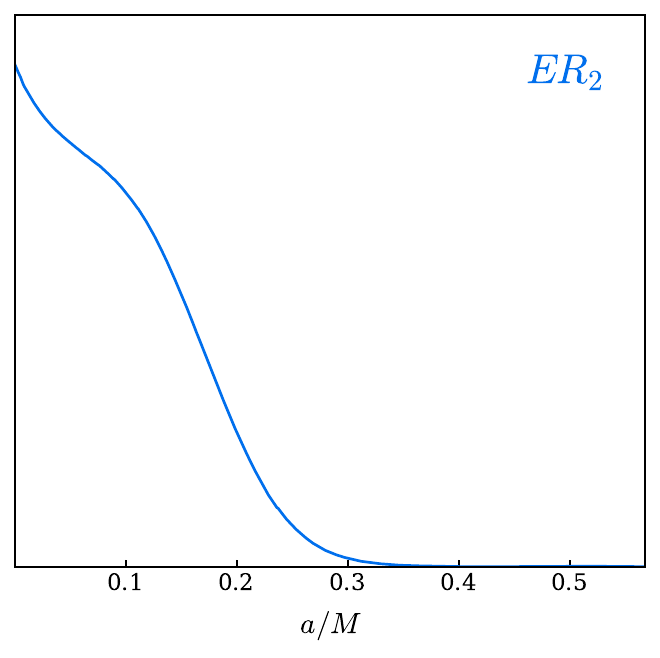}
\includegraphics[width=4.8cm]{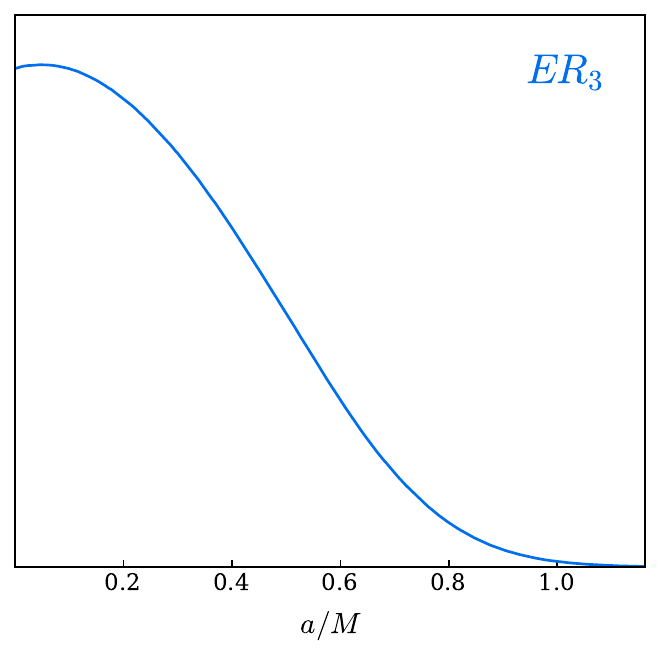}
\includegraphics[width=4.8cm]{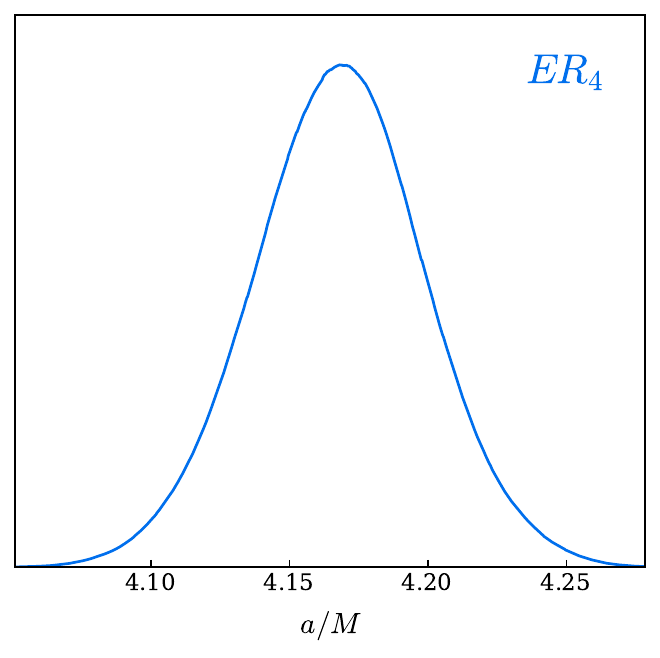}
\includegraphics[width=4.8cm]{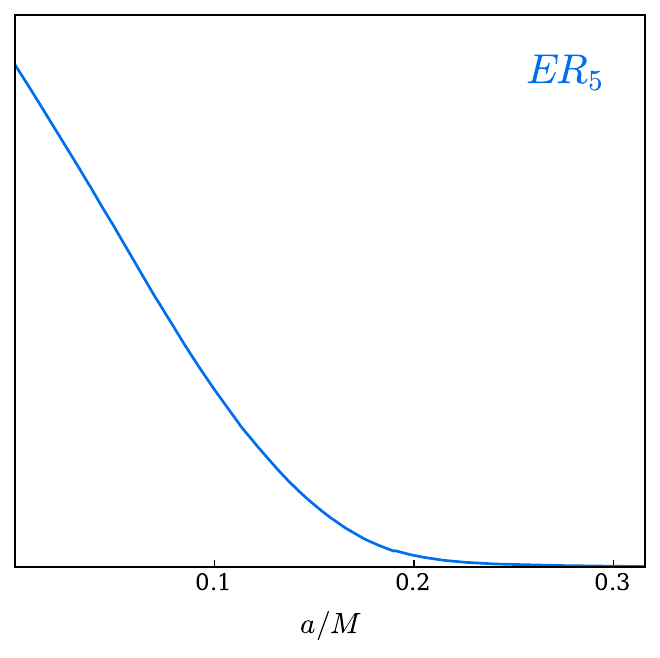}
\caption{For no magnetic case, the observational constraint on regularization parameter $a$ within the $68 \%$ and $95 \%$ CL from HFQPO data under different resonance models.}
    \label{fig:6}
\end{figure}

As a comparison, in this section, we use the same method as described in Section 6.1 to test the SV spacetime without a magnetic field. On basis of the observational  HFQPO data from the three microquasars, we construct the likelihood:
\begin{equation}
\chi^2=\sum_{k=1}^3 \frac{\left\{\nu_{u, k}-\nu_u\left(a/M, r_k^*\right)\right\}^2}{\sigma_{\nu_{u, k}}^2}+\sum_{k=1}^3 \frac{\left\{\nu_{l, k}-\nu_l\left(a/M, r_k^*\right)\right\}^2}{\sigma_{\nu_{l, k}}^2} . \label{40}
\end{equation}
Here, $r_k^*\;(k=1,2,3)$ represents the resonance positions around the three microquasars. It is important to note that in the absence of a magnetic field, we always have $\nu_\phi=\nu_\theta>\nu_r$. Therefore, the four cases of $E R_6-E R_9$ listed in Table \ref{table1} will not occur in the SV spacetime that is not immersed in a magnetic field. In Table \ref{table9}, we summarize the best-fit values of the regularization parameter $a/M$ within the $68 \%$ confidence interval and the $\chi_{\min }^2$ values for the different resonance models. The parameter plots for $a$ within the $68 \%$ and $95 \%$ confidence levels (CL) are shown in Figure \ref{fig:6}.  Finally, in Table \ref{table10}, we list the best-fit values of the circular orbit radii (resonance positions) for different models.
\begin{table}[ht]
\begin{center}
\begin{tabular}{|c|c|c|c|c|c|c|c|c|}
\hline \multirow{2}{*}{$E R_0$} & $a / M$ & $<0.175$ & \multirow{2}{*}{$E R_1$} & $a / M$ & $<0.233$ & \multirow{2}{*}{$E R_2$} & $a / M$ & $<0.129$ \\
\cline{2-3}\cline{5-6}\cline{8-9}& $\chi_{\text {min }}^2$ & 5055.43 & & $\chi_{\text {min }}^2$ & 810.46 & & $\chi_{\text {min }}^2$ & 7343.13 \\
\hline \multirow{2}{*}{$E R_3$} & $a / M$ & $<0.366$ & \multirow{2}{*}{$E R_4$} & $a / M$ & $4.163_{-0.039}^{+0.039}$ & \multirow{2}{*}{$E R_5$} & $a / M$ & $<0.0734$ \\
\cline{2-3}\cline{5-6}\cline{8-9} & $\chi_{\text {min }}^2$ & 27.98 & & $\chi_{\text {min }}^2$ & 58.49 & & $\chi_{\text {min }}^2$ & 31377.92 \\
\hline
\end{tabular}
\end{center}
\caption{\label{table9}
   The $68 \%$ confidence level of the regularization parameter $a$ and the minimum $\chi^2$ under different resonance models, for the SV spacetime that is not immersed in a magnetic field.}
\end{table}

\begin{table}[ht]
\begin{center}
\begin{tabular}{|l|l|l|l|l|l|l|}
\hline & \multicolumn{1}{|c|}{$E R_0$} & \multicolumn{1}{|c|}{$E R_1$} & \multicolumn{1}{|c|}{$E R_2$} & \multicolumn{1}{|c|}{$E R_3$} & \multicolumn{1}{|c|}{$E R_4$} & $E R_5$ \\
\hline$r_1^*$ & 6.319 & 6.098 & 6.020 & 7.207 & 4.880 & 7.560 \\
\hline$r_2^*$ & 6.339 & 6.077 & 6.025 & 6.879 & 4.762 & 7.420 \\
\hline$r_3^*$ & 6.643 & 6.387 & 6.131 & 8.183 & 5.707 & 7.467 \\
\hline
\end{tabular}
\end{center}
\caption{\label{table10}
   For no magnetic case, the best-fit values of the circular orbit radii associated with the three sets of HFQPOs under different resonance models.}
\end{table}

From Table \ref{table9}, it can be seen that for SV spacetime not immersed in a magnetic field, the resonance model with the smallest $\chi_{\min }^2$ is $\mathrm{ER}_3$. Compared to the best model shown in section 6.2, then we get $\triangle_i=23.357$ for  AIC (for BIC case, the value is the same). Since the models in Table \ref{table9} have the same number of free parameters, all these models will have $\triangle_i$ values greater than 23.357. This indicates that, under the resonance model, the observational data do not support SV spacetime without a magnetic field, suggesting that SV spacetime cannot simultaneously fit observational data from different sources. However, our study shows that considering the motion of charged particle around black holes surrounded by a magnetic field does help to resolve this issue, which further validates the importance of the magnetic field in SV spacetime.
\section{$\text{Conclusion}$}
Many observational facts suggest that magnetic fields likely exist around black holes. Due to the significant effect of Lorentz force on charged particle, even a weak magnetic field can greatly influence their trajectories, which in turn affects the observable properties of black holes, such as their shadow and the QPO phenomena. Therefore, when studying the dynamics of charged particle near black holes, it is crucial to consider the influence of external magnetic fields.

This paper investigates the dynamics of charged particle in magnetized SV spacetime, where we find that the presence of a magnetic field causes the ISCO of charged particle to move closer to the radial coordinate center. For magnetic fields of the same strength, the ISCO radius for $\mathcal{B}>0$ is consistently smaller than that for $\mathcal{B}<0$. Additionally, given that QPOs can serve as a powerful tool for testing gravitational theories, we study the oscillatory behavior of charged particle around the central object and calculate both epicyclic and orbital frequencies in magnetized SV spacetime. The results show that the magnetic field alters the distribution of frequencies, potentially exciting more variants of resonance model  and producing stronger observational effects.

The 3:2 twin-peak frequency ratios observed in the three microquasars GRS 1915+105, XTE 1550-564, and GRO 1655-40 indicate that nonlinear resonance mechanisms play a role in modulating X-ray flux. Here, we focus on discussing the resonance model and its variants, and examine the resonance locations in magnetized SV spacetime under different resonance models. The study shows that the presence of the magnetic field also shifts the resonance locations closer to the radial coordinate center.

Furthermore, through $\chi^2$ analysis, we use observational  HFQPO data from the three microquasars (GRS 1915+105, XTE 1550-564, and GRO 1655-40 to constrain the magnetized SV spacetime. Given that HFQPOs-related theoretical studies heavily depend on selected model, we evaluate the resonance model and its variants using the AIC. The AIC results show significant differences in the support for different models: $E R_8$ model is the best model and has received strong support, $ER_3$ model has received moderate evidence of support, and $ER_6$, $ER_7$ models have considerably less support, while other resonance models have essentially no support. The result from AIC is broadly consistent with the results obtained from the BIC and the Bayes factor.
For the models more supported by observations ($0 \leqslant \triangle_i$ $< 4$ ), the allowed ranges of the regularization parameter $0\leq a<0.736$ ($68\%$ CL), suggesting that the HFQPOs data could support the magnetized black hole bounce spacetime as a regular black hole. Additionally, the smaller value of the regularization parameter indicates a possibility of the presence of quantum effects. According to the constraint results, we get the best-fit values of magnetic field strength around $10^{-5}\sim 10^{-4}$ GS for electrons and around  $10^{-2}\sim 10^{-1}$ GS for protons.

Finally, as a comparison, we tested the SV spacetime without the presence of a magnetic field using observational data from the three microquasars. The results indicate that this model is essentially irrelevant to the observational data according to the calculated values of ICs and Bayesian analysis, suggesting that HFQPOs data support the existence of a magnetic field in SV spacetime.

\textbf{\ Acknowledgments }
 The research work is supported by the National Natural Science Foundation of China (12675070, 12175095), and supported by  LiaoNing Revitalization Talents Program (XLYC2007047).


\begin{thebibliography}{*}

\bibitem{1} B. P. Abbott et al., Phys. Rev. Lett. 116 (2016), 061102.
\bibitem{2} K. Akiyama et al., Astrophys. J. 875 (2019), L1.

\bibitem{wei} W. D. Guo, S. W. Wei, Y. X. Liu, Eur. Phys. J. C 83 (2023), 197.

\bibitem{3} K. Akiyama et al., Astrophys. J. 875 (2019), L4.
\bibitem{4} R. Torres,F. Fayos, Phys. Lett. B 733 (2014), 169-175.
\bibitem{5} R. Torres, Phys. Lett. B 733 (2014), 21-24.
\bibitem{6} M. R. Mbonye,D. Kazanas, IJPD17 (2008), 165-177.
\bibitem{7} K. Jusufifi, Universe 9 (2023), 41.
\bibitem{8} P. Bintruy,A. Helou,F. Lamy, Phys. Rev. D 98 (2018), 064058.
\bibitem{9} S. Riaz,S. Shashank,R. Roy,A. B. Abdikamalov et al., JCAP 10 (2022), 040.
\bibitem{10} M. Bojowald, Phys. Rev. Lett. 86 (2001), 5227-5230.
\bibitem{11} A. Ashtekar, T. Pawlowski, P. Singh, Phys. Rev. D 74 (2006), 084003.
\bibitem{12} A. Ashtekar, T. Pawlowski, P. Singh, Phys. Rev. Lett. 96 (2006), 141301.
\bibitem{13} Y. F. Cai, S. H. Chen, J. B. Dent, S. Dutta, E. N. Saridakis, Class. Quant. Grav. 28 (2011), 215011.
\bibitem{14} T. Qiu, J. Evslin, Y. F. Cai, M. Li, X. Zhang, JCAP 10 (2011), 036.
\bibitem{15} S. Alexander, C. Bambi, A. Marciano, L. Modesto, Phys. Rev. D 90 (2014), 123510.
\bibitem{16} J. M. Bardeen, Tbilisi, USSR, p. 174.
\bibitem{17} S. A. Hayward, Phys. Rev. Lett. 96(2006), 031103.
\bibitem{18} T. A. Roman, P. G. Bergmann, Phys. Rev. D 28 (1983), 12651277.
\bibitem{19} J. T. S. S. Junior,M. E. Rodrigues, Eur. Phys. J. C 83 (2023), 475.
\bibitem{20} V. P. Frolov, JHEP 1405 (2014), 049.
\bibitem{21} V. P. Frolov, Phys. Rev. D 94 (2016), 104056.
\bibitem{22} V. P. Frolov and A. Zelnikov, Phys. Rev. D 95 (2017), 124028.
\bibitem{23} R. Carballo-Rubio, F. Di Filippo, S. Liberati, C. Pacilio, M. Visser, JHEP 1807 (2018), 023.
\bibitem{24} R. Carballo-Rubio, F. Di Filippo, S. Liberati, M. Visser, Phy. Rev. D. 98 (2018), 124009.
\bibitem{25} A. Simpson, M. Visser, JCAP 02 (2019), 042.
\bibitem{26} Z. Stuchlk, J. Vrba, Universe 7 (2021), 279.16
\bibitem{27} J. Mazza, E. Franzin, S.Liberati, JCAP 04 (2021), 082.
\bibitem{28} E. Franzin, S. Liberati, J. Mazza, A. Simpson, M. Visser, JCAP 07 (2021), 036 .
\bibitem{29} N. Tsukamoto, Phys. Rev. D 104 (2021), 064022.
\bibitem{30} K. A. Bronnikov, R. A. Konoplya, T. D. Pappas, Phys. Rev. D 103 (2021), 124062.
\bibitem{31} R. Shaikh, K. Pal, T. Sarkar, Mon. Not. Roy. Astron. Soc. 506 (2021), 1229-1236.
\bibitem{32} N. Tsukamoto, Phys. Rev. D 103 (2021), 024033
\bibitem{33} F. S. N. Lobo, M. E. Rodrigues, M. V. d. S. Silva, A. Simpson, M. Visser, Phys. Rev. D 103 (2021), 084052
\bibitem{34} M. Guerrero, G. J. Olmo, D. Rubiera-Garcia, D. S. C. Gomez, JCAP 08 (2021), 036.
\bibitem{35} H. Huang, J. Yang, Phys. Rev. D 100 (2019), 124063.
\bibitem{36} Z. Xu, M. Tang, Eur. Phys. J. C 81 (2021), 863.
\bibitem{37} N. Chatzifotis, E. Papantonopoulos, C. Vlachos, Phys. Rev. D 105 (2022), 064025.
\bibitem{38} Y. Guo, Y. G. Miao, Nucl. Phys. B 983 (2022) 115938.
\bibitem{39} J. Barrientos, A. Cisterna, N. Mora, A. Vigan‘o, Phys. Rev. D 106 (2022), 024038.
\bibitem{40} S. U. Islam, J. Kumar, S. G. Ghosh, JCAP 10 (2021), 013.

\bibitem{42} X. Jiang, P. Wang, H. W. Wu, H. T. Yang, Eur. Phys. J. C,81 (2021),1043.


\bibitem{43} M. Guerrero,G. J. Olmo,D. Rubiera-Garcia,D. S. Gmez, Phys. Rev. D105 (2022), 084057.17
\bibitem{44} E. Franzin,S. Liberati,J. Mazza,R. Dey,S. Chakraborty, Phys. Rev. D 105 (2022), 124051.
\bibitem{45} L. Stella, M. Vietri, Phys. Rev. Lett. 82 (1999), 17.
\bibitem{46} Z. Stuchlik, A. Kotrlova, Gen. Relativ. Gravit. 41(2009), 1305.
\bibitem{47} A. Aliev, G. Esmer, P. Talazan, Class. Quantum Grav. 30 (2013), 045010.
\bibitem{48} T. Johannsen, D. Psaltis, Astrophys. J. 726 (2011), 11.
\bibitem{49} A. Maselli, L. Gualtieri, P. Pani, L. Stella, V. Ferrari, Astrophys. J. 801 (2015), 115.
\bibitem{50} F. Vincent, Class. Quantum Grav. 31 (2014), 025010.
\bibitem{51} A. Maselli, P. Pani, L. Gualtieri, V. Ferrari, Phys. Rev. D 92 (2015), 083014.
\bibitem{52} H. Wei, T. Jun, Z. TongJie, Phys. Rev. D 104 (2021) 124063.
\bibitem{53} J. Rayimbaev, K. F. Dialektopoulos, F. Sarikulov, Eur. Phys. J. C 83 (2023) 572.
\bibitem{54} A. Tursunov, Z. Stuchlik, M. Kolos, Phys. Rev. D 93 (2016) 084012.

\bibitem{56} K. L. Smith, C. R. Tandon, R. V. Wagoner, Astrophys. J. 906 (2021) 92.

\bibitem{57} J. Horak, M. Abramowicz, V. Karas, W. Kluzniak, Publ. Astron. Soc. Japan 56 (2004) 819--822.

\bibitem{58} J. Horak, arXiv:astro-ph/0408092.

\bibitem{59} P. Rebusco, Publ. Astron. Soc. Japan 56 (2004) 553.

\bibitem{60} G. T\"or\"ok, M. A. Abramowicz, W. Klu\'zniak, Z. Stuchl\'ik, Astron. Astrophys. 436 (2005) 1--8.

\bibitem{61} M. A. Abramowicz, W. Klu\'zniak, Astron. Astrophys. 374 (2001) L19--L20.

\bibitem{62} L. Stella, M. Vietri, Astrophys. J. Lett. 492 (1998) L59--L62.

\bibitem{63} L. Stella, M. Vietri, S. M. Morsink, Astrophys. J. 524 (1999) L63.

\bibitem{64} A. Cadez, M. Calvani, U. Kostic, Astron. Astrophys. 487 (2008) 527--532.

\bibitem{65} U. Kostic, A. Cadez, M. Calvani, A. Gomboc, Astron. Astrophys. 496 (2009) 307--315.

\bibitem{66} C. Germana, U. Kostic, A. Cadez, M. Calvani, AIP Conf. Proc. 1126 (2009) 367--369.

\bibitem{67} S. Kato, Publ. Astron. Soc. Japan 53 (2001) 1--24.

\bibitem{69} L. Rezzolla, S. Yoshida, T. J. Maccarone, O. Zanotti, Mon. Not. R. Astron. Soc. 344 (2003) L37--L41.

\bibitem{70} C. Bambi, J. Cosmol. Astropart. Phys. 09 (2012) 014.

\bibitem{72} A. Tursunov, Z. Stuchl\'ik, M. Kolo\v{s}, N. Dadhich, B. Ahmedov, Astrophys. J. 895 (2020) 14.

\bibitem{73} R. P\'anis, M. Kolo\v{s}, Z. Stuchl\'ik, Eur. Phys. J. C 79 (2019) 479, arXiv:1905.01186 [gr-qc].

\bibitem{74} S. Shaymatov, J. Vrba, D. Malafarina, et al., Phys. Dark Universe 30 (2020) 100648, arXiv:2005.12410 [gr-qc].


\bibitem{76} M. Kolo\v{s}, A. Tursunov, Z. Stuchl\'ik, Eur. Phys. J. C 77, 860 (2017).

\bibitem{77} R. M. Wald, Phys. Rev. D 10, 1680--1685 (1974).

\bibitem{78} S. Shaymatov, M. Jamil, K. Jusufi, K. Bamba, Eur. Phys. J. C 82, 636 (2022).

\bibitem{79} V. P. Frolov, A. A. Shoom, Phys. Rev. D 82, 084034 (2010), arXiv:1008.2985 [gr-qc].

\bibitem{80} A. N. Aliev, N. Ozdemir, Mon. Not. R. Astron. Soc. 336, 241 (2002), arXiv:gr-qc/0208025.

\bibitem{81} A. Abdujabbarov, B. Ahmedov, Phys. Rev. D 81, 044022 (2010), arXiv:0905.2730 [gr-qc].

\bibitem{82} S. Shaymatov, F. Atamurotov, B. Ahmedov, Astrophys. Space Sci. 350, 413 (2014).

\bibitem{83} M. Jamil, S. Hussain, B. Majeed, Eur. Phys. J. C 75, 24 (2015), arXiv:1404.7123 [gr-qc].

\bibitem{85} S. Hussain, M. Jamil, Phys. Rev. D 92, 043008 (2015), arXiv:1508.02123 [gr-qc].

\bibitem{86} M. Kolo\v{s}, Z. Stuchl\'ik, A. Tursunov, Class. Quantum Grav. 32, 165009 (2015), arXiv:1506.06799 [gr-qc].

\bibitem{87} S. Hussain, I. Hussain, M. Jamil, Eur. Phys. J. C 74, 210 (2014).

\bibitem{88} M. D. Laurentis, Z. Younsi, O. Porth, et al., Phys. Rev. D 97, 104024 (2018), arXiv:1712.00265 [gr-qc].

\bibitem{89} V. P. Frolov, P. Krtou\v{s}, Phys. Rev. D 83, 024016 (2011), arXiv:1010.2266 [hep-th].

\bibitem{90} V. P. Frolov, Phys. Rev. D 85, 024020 (2012), arXiv:1110.6274 [gr-qc].

\bibitem{91} S. Shaymatov, M. Patil, B. Ahmedov, P. S. Joshi, Phys. Rev. D 91, 064025 (2015), arXiv:1409.3018 [gr-qc].

\bibitem{92} B. Narzilloev, J. Rayimbaev, S. Shaymatov, et al., Phys. Rev. D 102, 044013 (2020), arXiv:2007.12462 [gr-qc].

\bibitem{93} S. Shaymatov, D. Malafarina, B. Ahmedov, Phys. Dark Universe 34, 100891 (2021), arXiv:2004.06811 [gr-qc].

\bibitem{94} B. Narzilloev, J. Rayimbaev, S. Shaymatov, et al., Phys. Rev. D 102, 104062 (2020), arXiv:2011.06148 [gr-qc].

\bibitem{95} S. Shaymatov, B. Ahmedov, M. Jamil, Eur. Phys. J. C 81, 588 (2021).

\bibitem{96} A. R. Prasanna, V. Vishveshwara, Pramana 11, 359 (1978).

\bibitem{97} A. R. Prasanna, Nuovo Cimento Rivista Serie 3, 1 (1980).

\bibitem{98} A. N. Aliev, D. V. Galtsov, Gen. Relativ. Gravit. 13, 899 (1981).

\bibitem{99} A. A. Abdujabbarov, B. J. Ahmedov, N. B. Jurayeva, Phys. Rev. D 87, 064042 (2013).

\bibitem{100} A. M. Al Zahrani, V. P. Frolov, A. A. Shoom, Phys. Rev. D 87, 084043 (2013), arXiv:1301.4633.

\bibitem{102} O. Kop\'a\v{c}ek, V. Karas, Astrophys. J. 787, 117 (2014), arXiv:1404.5495.

\bibitem{103} R. Shiose, M. Kimura, T. Chiba, Phys. Rev. D 90, 124016 (2014), arXiv:1409.3310.

\bibitem{105} O. Kop\'a\v{c}ek, V. Karas, Astrophys. J. 853, 53 (2018), arXiv:1801.01576.

\bibitem{106} C. A. Benavides-Gallego, A. Abdujabbarov, D. Malafarina, B. Ahmedov, C. Bambi, Phys. Rev. D 99, 044012 (2019).

\bibitem{108} K. A. Bronnikov, R. K. Walia, Phys. Rev. D 105, 044039 (2022), arXiv:2112.13198.

\bibitem{109} C. M. Hull, JHEP 11, 017 (1998).

\bibitem{110} T. Okuda, T. Takayanagi, JHEP 03, 062 (2006).

\bibitem{111} H. G. Ellis, J. Math. Phys. 14, 104 (1973).

\bibitem{112} K. A. Bronnikov, Acta Phys. Pol. B 4, 251 (1973).

\bibitem{113} T. Karakasis, E. Papantonopoulos, C. Vlachos, Phys. Rev. D 105, 024006 (2022), arXiv:2107.09713.

\bibitem{114} K. A. Bronnikov, Phys. Rev. D 106, 064029 (2022), arXiv:2206.09227.

\bibitem{115} X. Qin, S. B. Chen, J. L. Jing, Class. Quantum Grav. 38, 115008 (2021).

\bibitem{116} K. A. Bronnikov, J. C. Fabris, Phys. Rev. Lett. 96, 251101 (2006).

\bibitem{117} G. W. Gibbons, D. A. Rasheed, Nucl. Phys. B 476, 515 (1996), arXiv:hep-th/9604177.

\bibitem{118} G. Cl\'ement, J. C. Fabris, M. E. Rodrigues, Phys. Rev. D 79, 064021 (2009), arXiv:0901.4543.

\bibitem{119} M. Azreg-A\"inou, G. Cl\'ement, J. C. Fabris, M. E. Rodrigues, Phys. Rev. D 83, 124001 (2011), arXiv:1102.4093.

\bibitem{120} C. J. Gao, S. N. Zhang, Phantom Black Holes, arXiv:hep-th/0604114.

\bibitem{121} M. E. Rodrigues, Z. A. A. Oporto, Phys. Rev. D 85, 104022 (2012), arXiv:1201.5337.

\bibitem{122} D. F. Jardim, M. E. Rodrigues, M. J. S. Houndjo, Eur. Phys. J. Plus 127, 123 (2012), arXiv:1202.2830.

\bibitem{123} A. Nakonieczna, M. Rogatko, R. Moderski, Phys. Rev. D 86, 044043 (2012), arXiv:1209.1203.

\bibitem{124} S. V. Bolokhov, K. A. Bronnikov, M. V. Skvortsova, Class. Quantum Grav. 29, 245006 (2012).

\bibitem{125} S. Chen, J. Jing, Class. Quantum Grav. 30, 175012 (2013).



\bibitem{151} W. Godlowski, M. Szydlowski, Phys. Lett. B 623, 10 (2005), arXiv:astro-ph/0507322.

\bibitem{PlanckCollaboration2020} N. Aghanim et al. [Planck Collaboration], Astron. Astrophys. 641, A6 (2020), arXiv:1807.06209.

\bibitem{DESI2025} A. G. Adame et al. [DESI Collaboration], JCAP 02, 021 (2025), arXiv:2404.03002.

\bibitem{Caldwell2002} R. R. Caldwell, Phys. Lett. B 545, 23 (2002), arXiv:astro-ph/9908168.

\bibitem{Carroll2003} S. M. Carroll, M. Hoffman, M. Trodden, Phys. Rev. D 68, 023509 (2003), arXiv:astro-ph/0301273.

\bibitem{126} R. Javlon, D. Alexandra, U. Camci, A. Abdujabbarov, B. Ahmedov, arXiv:2010.04969 [gr-qc].

\bibitem{127} R. M. Wald, Ann. Phys. 82, 548 (1974).

\bibitem{128} R. M. Wald, \textit{General Relativity} (University of Chicago Press, Chicago, 1984).

\bibitem{131} C. A. Benavides-Gallego, A. Abdujabbarov, D. Malafarina, C. Bambi, Phys. Rev. D 101, 124024 (2020).

\bibitem{132} I. Banerjee, JCAP 08, 034 (2022).

\bibitem{133} W. Klu\'zniak, M. A. Abramowicz, arXiv:astro-ph/0203314.

\bibitem{135} W. Klu\'zniak, M. A. Abramowicz, Acta Phys. Pol. B 32, 3605 (2001).

\bibitem{136} Z. Stuchl\'ik, M. Kolo\v{s}, Astron. Astrophys. 586, A130 (2016).

\bibitem{137} I. Banerjee, arXiv:2203.10890.

\bibitem{139} J. Horak, M. Abramowicz, V. Karas, W. Klu\'zniak, Publ. Astron. Soc. Japan 56, 819 (2004).

\bibitem{140} M. A. Abramowicz, V. Karas, W. Klu\'zniak, et al., Publ. Astron. Soc. Japan 55, 466--467 (2003), arXiv:astro-ph/0302183.

\bibitem{142} L. D. Landau, E. M. Lifshitz, \textit{Mechanics} (Pergamon Press, 1969).

\bibitem{143} E. Deligianni, J. Kunz, P. Nedkova, S. Yazadjiev, R. Zheleva, Phys. Rev. D 104, 024048 (2021).

\bibitem{144} GRAVITY Collaboration, R. Abuter et al., Astron. Astrophys. 618, L10 (2018).

\bibitem{145} K. V. Staykov, D. Popchev, D. D. Doneva, S. S. Yazadjiev, Eur. Phys. J. C 78, 586 (2018).

\bibitem{146} G. T\"or\"ok, A. Kotrlov\'a, E. \v{S}r\'amkov\'a, Z. Stuchl\'ik, Astron. Astrophys. 531, A59 (2011).

\bibitem{147} M. E. Beer, P. Podsiadlowski, Mon. Not. R. Astron. Soc. 331, 351 (2002), arXiv:astro-ph/0109136.

\bibitem{148} J. A. Orosz, J. F. Steiner, J. E. McClintock, et al., Astrophys. J. 730, 75 (2011), arXiv:1101.2499 [astro-ph.SR].

\bibitem{149} M. J. Reid, J. E. McClintock, J. F. Steiner, et al., Astrophys. J. 796, 2 (2014), arXiv:1409.2453 [astro-ph.GA].

\bibitem{150} A. R. Liddle, Mon. Not. R. Astron. Soc. 351, L49 (2004), arXiv:astro-ph/0401198.

\bibitem{152} M. Szydlowski, W. Godlowski, Phys. Lett. B 633, 427 (2006), arXiv:astro-ph/0509415.

\bibitem{153} J. B. Lu, L. X. Xu, J. C. Li, B. R. Chang, Y. X. Gui, H. Y. Liu, Phys. Lett. B 662, 87--91 (2008).

\bibitem{154} M. Biesiada, J. Cosmol. Astropart. Phys. 02, 003 (2007), arXiv:astro-ph/0701721.

\bibitem{157} L. X. Xu, C. W. Zhang, H. Y. Liu, Chin. Phys. Lett. 24, 2459 (2007).

\bibitem{158} M. Szydlowski, A. Kurek, A. Krawiec, Phys. Lett. B 642, 171 (2006), arXiv:astro-ph/0604327.

\bibitem{BIC} G. Schwarz, Ann. Stat. 6, 461--464 (1978).

\bibitem{Bys-1} Y. L. Kong, J. D. Zhang, Phys. Rev. D 110, 024059 (2024).

\bibitem{Bys-2} E. Thrane, C. Talbot, Publ. Astron. Soc. Aust. 36, e010 (2019).

\bibitem{Bys-3} G. H. Du, P. J. Wu, T. N. Li, X. Zhang, Eur. Phys. J. C 85, 392 (2025).

\bibitem{Bys-4} R. E. Kass, A. E. Raftery, J. Am. Stat. Assoc. 90, 773--795 (1995).

\bibitem{Bys-5} R. Trotta, Contemp. Phys. 49, 71--104 (2008).

\bibitem{159} M. Qi, J. Rayimbaev, B. Ahmedov, Eur. Phys. J. C 83, 730 (2023).
\end{thebibliography}
\end{document}